\documentclass{emulateapj}

\newcommand{\eg}{{e.g.,}}

\shorttitle{Intermediate-Age Populations in QSO Hosts}
\shortauthors{Canalizo \& Stockton}

\begin{document}

\title{Intermediate-Age Stellar Populations in Classical QSO Host Galaxies}

\author{Gabriela Canalizo}
\affil{Department of Physics and Astronomy, University of California, Riverside, CA 92521: gabriela.canalizo@ucr.edu}

\and

\author{Alan Stockton} 
\affil{Institute for Astronomy, University of Hawaii, 2680 Woodlawn Dr., Honolulu, HI 96822: stockton@ifa.hawaii.edu}

\begin{abstract}
Although mergers and starbursts are often invoked in the discussion of QSO activity in the context of galaxy evolution, several studies have questioned their importance or even their presence in QSO host galaxies. Accordingly, we are conducting a study of $z\sim0.2$ QSO host galaxies previously classified as passively evolving elliptical galaxies. We present deep Keck LRIS spectroscopy of a sample of 15 hosts and model their stellar absorption spectra using stellar synthesis models. The high S/N of our spectra allow us to break various degeneracies that arise from different combinations of models, varying metallicities, and contamination from QSO light. We find that none of the host spectra can be modeled by purely old stellar populations and that the majority of the hosts (14/15) have a substantial contribution from intermediate-age populations with ages ranging from 0.7 to 2.4 Gyr. An average host spectrum is strikingly well fit by a combination of an old population and a 2.1 (+0.5, $-$0.7) Gyr population. The morphologies of the host galaxies suggest that these aging starbursts were induced during the early stages of the mergers that resulted in the elliptical-shaped galaxies that we observe. The current AGN activity likely corresponds to the late episodes of accretion predicted by numerical simulations, which occur near the end of the mergers, whereas earlier episodes may be more difficult to observe due to obscuration. Our off-axis observations prevent us from detecting any current star formation or young stellar populations that may be present in the central few kiloparsecs.
\end{abstract}


\section{Introduction}

A paradigm is beginning to emerge that links galaxy mergers, starbursts, QSOs, and the formation of spheroids in galaxies into a single coherent picture that promises to explain a range of disparate
phenomena connected to galaxy formation and evolution (\eg\ \citealt{hopkins2006a}).  In outline, galaxy
mergers involving gas-rich galaxies drive gas into the center of the resulting merger remnant, 
fueling both a starburst and the black hole that results from the merger of the central black holes of
the merging galaxies. Initially, both the resulting QSO activity and much of the starburst are hidden
at optical wavelengths by enshrouding dust, but eventually the strong outflow from the QSO clears
a path along at least some sightlines \citep{canalizo2001,dimatteo2005}. This same outflow, however, also clears
out much of the gas that is feeding both the starburst and the QSO activity, quenching both within a
relatively short time and providing a self-limiting mechanism for the growth of both the black hole 
and the stellar mass of the galaxy.  This removal of the gas feeding the black hole also means
that the optically luminous phase of the QSO is necessarily brief, $\sim10^7$--$10^8$ years.
The coupling between black-hole growth and star formation
provides a plausible explanation for the correlation between black-hole mass and spheroidal
velocity dispersion \citep{geb00a,fm2000}, and relations between the QSO luminosity and the strength of the outflow may also explain the shape of the QSO luminosity function \citep{hopkins2005a}.  Furthermore, 
there may be a link between the QSO and merging galaxy luminosity functions \citep{hopkins2006b}.

This close relationship between gas-rich mergers and nuclear activity is supported by evidence for recent starbursts in galaxies with high-luminosity AGN \citep{canalizo2001,sanchez2004,letawe2007,jahnke2007,wold2010}. However, it is seemingly difficult
to reconcile such star formation with claims that most QSOs in the low-redshift universe are found in normal ellipticals with
stellar populations overwhelmingly dominated by old stars \citep{mclure1999,nolan2001,dunlop2003}.  Indeed, much skepticism remains concerning the relevance of major mergers (and their ensuing starbursts) in the triggering of AGN activity \citep[e.g.,][]{cisternas2011,kocevski2012}, although many agree that mergers do play an important role for higher luminosity AGN, such as QSOs \citep[e.g.,][]{treister2012}.

Given that determining the relationship between mergers and the QSO population has important consequences in our understanding of galaxy evolution, it is pertinent to revisit the question whether or not QSO hosts have populations indicative of merger-induced starbursts.
\citet{bennert2008} studied the morphologies of five low-redshift QSO host galaxies that had been previously classified as undisturbed elliptical galaxies using deep (5 orbit) Hubble Space Telescope ($HST$) Advanced Camera for Surveys (ACS) images.   They found that, while the hosts are reasonably fit by a de Vaucouleurs profile, they also show significant tidal debris indicative of a merger event within the last $\sim2$ Gyr.   

Following \citet{bennert2008}, we now turn to the stellar populations in these and other similar bulge-dominated QSO host galaxies.    Using high signal-to-noise ratio (S/N), deep Keck spectroscopy, we model the stellar populations to search for traces of significant starburst episodes in the recent past.   In Section~\ref{observations} we describe the sample, spectroscopic observations, and data reduction.  In Section~\ref{modeling} we describe our modeling strategy and consider possible caveats, while in Section~\ref{compare} we compare our results to those of similar studies.   In Section~\ref{discussion} we discuss the interpretation of our results in the framework of galaxy evolution.    Throughout the paper, we assume a cosmological model with $H_{0} = 71$ km\,s$^{-1}$Mpc$^{-1}$, $\Omega_m=0.27$ and $\Omega_{\Lambda} = 0.73$ \citep{spergel2003}.

\section{Observations and Data Reduction}\label{observations}

\subsection{Sample Selection}\label{sample}

Our sample is drawn from the sample of 23 (10 radio-loud and 13 radio-quiet) QSOs of \citet{dunlop2003}.   This sample contains radio-loud and radio-quiet QSOs in the redshift range of $0.1<z<0.26$ that are matched in optical luminosity.  For details of the original sample selection, see \citet{dunlop1993}.  

\citet{dunlop2003} and \citet{mclure1999} performed a morphological study of this sample based on $HST$ images and concluded that all of the host galaxies having nuclei in the QSO luminosity range are bulges that have properties ``indistinguishable from those of quiescent, evolved, low-redshift ellipticals of comparable mass''  \citep{dunlop2003}.    Similarly, by studying the stellar populations using spectroscopy obtained with 4-m class telescopes \citep{hughes2000}, \citet{nolan2001}  conclude that these galaxies are dominated by truly old stellar populations with no significant episodes of star formation in the more recent past.

For this paper, we obtained spectra of as many of the 23 QSOs as we were able to observe during our runs, giving priority to those objects classified as ellipticals by \citet{dunlop2003}.   The 17 targets that we observed are listed in Table~\ref{journal}, along with their host galaxy redshifts as measured from absorption lines in our spectra.   We use the IAU designation listed in this table to refer to any given object throughout the paper.

\begin{deluxetable*}{llclrcc}
\tablecaption{Journal of Spectroscopic Observations \label{journal}}
\tablehead{\colhead{IAU} &Host & \colhead{Slit PA} &\multicolumn{2}{c}{Slit Offset} & \colhead{Dispersion} & \colhead{Total Int.} \\
\colhead{Designation} & $z_{\rm abs}$ &\colhead{(deg)} & \colhead{(\arcsec)} & \colhead{(kpc)}
& \colhead{(\AA\ pixel$^{-1}$)} & \colhead{Time (s)} }
\startdata
0054+144 &0.1721&\phn32.0& 2.5 SE &7.2& 0.63, 1.86  &   10800\phn\\
0137+012 &0.2640& 341.3  & 3.0 E  &12.1 & 0.63, 1.86  &    5400\\
0157+001 &0.1637&103.0  & 4.0 N    &11.1 & 2.44        &    3600\\
0204+292 &0.1102&\phn\phn0.0 & 3.0 E   &6.0 & 0.63, 1.86  &    1200\\
0244+194 &0.1745&\phn90.0& 2.6 N    &7.6 & 0.63, 1.86  &    5400\\
0736+017 &0.1886& 351.5  & 3.0 W    &9.4& 1.09, 2.55  &   10800\phn\\
0923+201 &0.1935&\phn70.3& 2.5 NNE  &8.0 & 1.09, 1.86  &    7200\\
1004+130 &0.2415&\phn70.0& 2.5 ESE  &9.4 & 1.09, 1.86  &    7200\\
1012+008 &0.1856& $-$61.4& 0.0      &0.0 & 2.44        &    1200\\
1020$-$103&0.1957& 341.6 & 2.5 NW   &8.0 & 1.09, 2.55  &    7200\\
1217+023 &0.2408& 337.0  & 2.5 WSW  &9.4 & 1.09, 1.86  &    7200\\
1549+203&0.2527$^{a}$&\phn72.0 & 2.0 NNW  &7.8 & 1.09, 1.86 &    7200\\
1635+119 &0.1476&\phn57.0& 0.0      &0.0 & 1.09, 2.55  &    5400\\
2135$-$147&0.1994& 343.8 & 3.0 WSW  &9.8 & 0.63, 1.86  &    5400\\
2141+175 &0.2115&\phn90.0& 2.2 N    &7.5& 0.63, 1.86  &    5400\\
2247+140 &0.2340&\phn40.0& 2.5 NE   &9.2 & 0.63, 1.86  &    5400\\
2349$-$014&0.1746&\phn\phn0.0&3.0 W &8.8 & 0.63, 1.86  &    5400\\
\enddata
\tablecomments{$^{a}$Redshift for 1549+203 measured from narrow emission lines.}
\end{deluxetable*}

\subsection{Spectroscopic Observations}\label{spectroscopy}
Spectroscopic observations were carried out with the Low-Resolution 
Imaging Spectrometer \citep[LRIS;][]{oke1995} on the Keck I telescope on the nights of  UT 2002 March 4 and 8, and 2002 October 3 and 12.
For the blue side (LRIS-B), we used either the 400 groove mm$^{-1}$ grism 
blazed at 3400\,\AA\ or the 600 groove mm$^{-1}$ grism blazed at 4000\,\AA\ 
yielding dispersions of 1.09\,\AA\ pixel$^{-1}$ and 0.63\,\AA\ pixel$^{-1}$,
respectively. For the red side (LRIS-R), we used either the 300 groove mm$^{-1}$ grating 
blazed at 5000\,\AA\ or the 400 groove mm$^{-1}$ grating blazed at 8500\,\AA,
yielding dispersions of 2.55~\AA\ pixel$^{-1}$ and 1.86~\AA\ respectively.
The slit was 1\arcsec\ wide, projecting to $\sim$7 pixels on UV- and blue-optimized
CCD on LRIS-B and $\sim$5 pixels on the Tektronix 2048$\times$2048 CCD
on LRIS-R.  A few of the objects were observed with a set up that resulted in a gap
of a few hundred angstroms between the blue and red sides.
We obtained between two and six exposures for each host galaxy, 
typically 1800~s each, dithering along the slit between exposures.
A spectrum of the QSO nucleus for each object was obtained separately, 
with total integration times between 600 and 1200~s, depending on the 
magnitude of each QSO, so as to obtain a S/N $\geq$ 100.
The typical seeing for all observations was 0\farcs7 in $B$.
Four or five spectrophotometric standards from \citet{massey1988} were observed 
each night for flux calibration.

Objects 0157+001 and 1012+008 were observed with the old LRIS (single camera) on the Keck~II 
telescope on UT 1997 September 13 and 1998 April 6, respectively.  We used a 300 groove mm$^{-1}$ grating blazed at 5000\,\AA\ with
a dispersion of 2.44~\AA\ pixel$^{-1}$. The slit was also 1\arcsec\ wide, 
projecting to $\sim$5 pixels on the Tektronix 2048$\times$2048 CCD.   

In Table \ref{journal} we show a complete journal of observations, 
including slit positions, total exposure times, and wavelength resolutions for the blue and red sides, respectively.   

The slit was placed offset from the nucleus for each object, with a separation that depended on the brightness of the QSO and the seeing conditions, so as to minimize the contamination from QSO light.   Thus, the spectra that we obtained correspond to regions of the host galaxy at a given distance from the center, as listed in Table~\ref{journal}.  For two objects, 1012+008 and 1635+119, we placed the slit through the nucleus.  However, the spectra that we recovered corresponded to regions at a distance $>$5 kpc from the nucleus.

The majority of objects were observed near transit to minimize the effects of 
differential atmospheric refraction.  In a few cases when the objects were 
observed at somewhat higher airmasses (1.1 to 1.4), shorter exposures at
parallactic angle were obtained to correct the continuum of the QSO and the 
host galaxy.  Only one object, 0204+292, was observed at high airmass ($\sim$2).

Spectra were reduced using standard procedures and corrected for galactic 
reddening using the values given by NED, which are calculated following 
\citet{schlegel1998}.   For each object, a scaled version of the QSO spectrum was 
subtracted from that of the host galaxy.  In order to model the contribution 
of the QSO to the host galaxy spectrum at each wavelength, we obtained spectra
of bright standard stars by placing the slit at several distances from the
star and at different angles to simulate our off-nuclear observations of the
host galaxies.  We then multiplied the spectrum of each QSO by the 
corresponding model. Finally, we scaled the model of the QSO contribution for 
each object by measuring the flux in broad lines in the spectrum of the host
and subtracted it from the latter.   The precise determination of the scaling of the QSO spectrum
was an iterative process where we tested a range of factors until we obtained clearly over- and under-subtracted host spectra.  We used these extreme cases to determine the effects of the uncertainty introduced by
the subtraction of the QSO light from the hosts on the modeling of the stellar populations, as described below.  The QSO contribution to the observed
host spectrum that was subtracted, as measured at rest wavelength 4500 \AA, is listed as a percentage value in Table~\ref{ages}.

The red-side spectrum of 0244+194 was corrupted, so we only used the blue-side spectrum when modeling stellar populations.
As mentioned above, 0204+292 was observed at high airmass and the host galaxy spectrum suffered from strong contamination from the QSO that we were not able to model appropriately.  Therefore, we were unable to recover a clean spectrum of the host galaxy.   For this reason, this object was not included in the analysis.   The spectrum of the host galaxy of 1549+023 was also 
too noisy to be modeled and it is also excluded from the analysis.   Our final sample thus consists of 15 objects.

\section{Modeling of Stellar Populations}\label{modeling}

\subsection{Fitting method}\label{method}
Several different strategies for modeling stellar populations in QSO and other AGN host galaxies have been developed over the last two decades.   Some authors \citep[e.g.,][]{sanchez2004,jahnke2004} use the SEDs and colors of hosts obtained from images, while others \citep[e.g.,][]{kauffmann2003} use diagnostic diagrams such as the $D_{n}$(4000)/H$\delta_A$ plane.   Although these techniques are powerful to analyze large samples of galaxies, they could potentially be vulnerable to degeneracies, such as the age-metallicity degeneracy, and contamination by scattered QSO light \citep[e.g.,][]{young2009}.

There are relatively few studies in the literature that include modeling of absorption-line spectroscopy of QSO hosts.   Even in those few studies, the techniques employed are vastly different.  For example, \citet{nolan2001} model off-nuclear spectra of QSO hosts in the \citet{dunlop2003} sample using stellar synthesis models.   They use the nuclear spectrum of one of the QSOs in the sample to account for QSO light contamination, and they fix a 0.1 Gyr population to account for any recent star formation that may have occurred.   They then vary the age of the old underlying population and determine the relative contribution between the older and the younger populations.   One problem with this method is that the spectra of younger populations change much more rapidly with age than those of older populations.  Whereas the SEDs and stellar features change vastly from 0 to 1 Gyr, they remain almost unchanged for ages greater than 6 Gyr, and the only parameter that changes significantly is the contribution by mass to the spectrum.   Thus, this method is prone to result in more degeneracies, and it is unable to determine the actual age and contribution of the younger population.  Not surprisingly, they find that the host galaxies are dominated by old stellar populations with a very small contribution of the 0.1 Gyr population

\citet{wold2010} do a more detailed modeling of the off-nuclear spectra of ten QSOs.    Their fitting method is based on that of \citet{cidfernandes2005}: They fit simultaneously instantaneous burst models of 15 different ages ranging from 0.001 to 13 Gyr.  In addition, they account for scattered QSO light and reddening.   In contrast to the results by \citet{nolan2001}, they find an average light-weighted age of 2 Gyr for the host galaxies in their sample. 

Our approach is a simplified version of the technique used by \citet{wold2010}.  The main question we want to answer is whether the host galaxies were formed at high redshift, with their populations only passively evolving, or whether there have been any significant episodes of star formation in the more recent past.   Therefore, we have chosen to employ  a method that we have used in the past \citep{canalizo2007,canalizo2001,canalizo2000a,canalizo2000b} which has proved to be very robust.   We assume an old stellar population (with a fixed age) corresponding to the populations existing in the host galaxy or galaxies.   To this, we add a younger stellar population representing a star forming event in the more recent past.   We then vary the relative contribution of the two stellar populations and the age of the younger population, and we perform a least square fit to the observed data, masking out regions of obvious emission lines and any regions showing excessive noise, and giving a lower relative weight to regions that contained pure stellar continuum, i.e., no absorption features.    A clear advantage of this method is that there are only two free parameters and thus we encounter fewer degeneracies.   

We typically assume the old population to be a 10 Gyr-old population, although we have found that the precise choice of the older population has no significant effect on the age of the younger population determined from the best fit.  However, the choice of the old population does have an effect on the determination of the relative contribution between the older and younger populations, in the sense that over-estimating the age of the older population will result into an over-estimation of the contribution of the younger population and vice versa.  This systematic uncertainty is of the order of a few percent, even when the age of the old population is varied by as much as five Gyr.

We use the stellar population synthesis models by \citet[][heareafter BC03]{bc2003} because they provide a good match to the spectral resolution of our data and allow us to test the effects of varying a wide range of metallicities. They are also widely used in the literature and allow for us to readily compare our results to those of previous studies.   While a recent study by \citet{zibetti2013} indicates that the BC03 models are still the most successful at reproducing the observed SEDs and features of post-starburst galaxies, we have chosen to also use two other sets of models that include different AGB star prescriptions: those by \citet[][hereafter M05]{maraston2005} and S.\ Charlot \& G.\ Bruzual (2007, private communication; hereafter CB07).  As we discuss below, we find remarkable consistency in the results obtained with the three sets of models.  The description of the fitting and analysis below refers to the BC03 models, unless otherwise specified.

In carrying out the fitting procedure, we smoothed the models to match the velocity dispersion of the host galaxies and the various resolutions of the spectra. Velocity dispersions were determined by using the penalized pixel-fitting procedure (pPXF) of \citet{cappellari2004}. We followed their prescription for adjusting the bias parameter, although, for our spectra, the results were not particularly sensitive to this choice. 

 Although the uncertainties in the determination of the QSO contribution to the host spectrum (listed in Table~\ref{ages}) are only of the order of a few percent, they could have an effect in the determination of the age of the stellar populations.   In order to estimate the magnitude of this effect, we created versions of host spectra in which the QSO contribution was clearly under- and over-subtracted, and we modeled these extreme spectra.   The best-fit stellar population ages determined for the under- and over-subtracted spectra were somewhat different from those determined for the spectra with the best QSO subtractions, but the difference was typically smaller than the 68\% confidence intervals in the fits given in Table~\ref{ages}.  Moreover, the $\chi^{2}$ of the best fits in the extreme spectra was significantly greater.

Many of the spectra of the individual objects show narrow emission lines.   In most cases, the emission lines come from gas ionized by the QSO in the extended narrow line region (NLR).   However, in a few cases (discussed in more detail in the Appendix), we see [\ion{O}{2}] $\lambda$3727 even when [\ion{O}{3}] is weak or absent.   This may be indicative of a small amount of current star formation or low velocity shocks in these hosts.

The presence of narrow emission lines in the host galaxy spectra affects the strength of the observed Balmer absorption features.   In fact, Balmer absorption features cannot be relied on because of (1) uncertainty in the ratio of narrow to broad emission in the scaled QSO spectrum relative to that in the scattered light (i.e., the QSO spectrum may miss some of the NLR, which will still be present in the scattered component), and (2) the frequent direct contamination from extended narrow-line emission. For this reason, we do not use their equivalent widths or their corresponding Lick indices directly in our analysis. The Balmer absorption lines, when available, do still enter as one component in our spectral fits. In some cases, the emission is clearly narrower than the absorption and can be masked out during the fitting procedure.

For any objects for which there were concerns about the accuracy of the spectrophotometric calibration, we normalized the continuum of both the host galaxy spectrum and the models to unity to eliminate any dependance on the shape of the continuum. In all cases, these fits gave essentially the same starburst ages as the original fits that included the continuum shape.

As an additional check, we have attempted to analyze the stellar populations, again using the pPXF routine of \citet{cappellari2004}. In the mode we used, pPXF finds the best linear combination of spectral synthesis models to fit the observed spectrum over the rest-frame 3500--4500 \AA. A 4th-order multiplicative polynomial is included in the fit to remove the low-order dependence on flux calibration and reddening. Because this procedure introduces many more free parameters, it will work best for the spectra for which we have the highest S/N. This indeed seems to be the case:  for the roughly half of the sample with the highest S/N and the least impact from extended emission, we find significant contributions from populations with ages within the range given in Table 2. For all but one of the remaining QSO hosts, we find contributions from intermediate-age populations with ages slightly outside this range. For 2135$-$147, this procedure finds only an old stellar population; but this host galaxy is the one most strongly impacted by an extended emission region, so Balmer absorption lines have no influence on the solution.

We used models with metallicities ranging from 0.004 to $2.5\,Z_{\sun}$ to fit the host galaxy spectra.   Invariably, models with $Z < 0.4\,Z_{\sun}$ led to poor fits or did not converge.   For most objects, the fits with the lowest $\chi^{2}$ values corresponded to those of models with solar metallicity.
Even in cases where models with metallicities somewhat lower or higher than solar gave reasonable fits, the ages of the best fitting populations changed only by $\sim30$\%.  For consistency, we report results using solar metallicity models for all objects, except for 1217+023, where a $2.5\,Z_{\sun}$ resulted in a significantly better fit.   See Section~\ref{oldpop} for further discussion on metallicities.

\subsection{Results: Starburst Ages}\label{results}

Our results are summarized in Figure~\ref{figure:bestfit}  and Table~\ref{ages}.   Figure~\ref{figure:bestfit} shows each of the QSO host spectra in the sample along with the best-fit model.   As mentioned above, each model consists of a 10 Gyr old population plus a younger (typically of intermediate-age) starburst.  The models have been matched in resolution to the host-galaxy spectra, taking into account both the instrumental profile and the velocity dispersion of the host galaxy.  In Table~\ref{ages}, we list the relative contribution at 4500 \AA\ (rest wavelength) of the QSO component that was subtracted from the observed host spectrum as described above, and the off-nuclear stellar velocity dispersion, $\sigma_{v}$. We also list the starburst ages and the percentage by mass that they contribute to the best-fit models.   The age range given corresponds to the 68\% confidence intervals in the fits for starburst ages, with their respective contributions listed as a contribution range.  The first striking result is that the majority of the hosts have a substantial contribution from an intermediate-age starburst population.

\begin{figure*}[p]
\epsscale{1.0}
\plotone{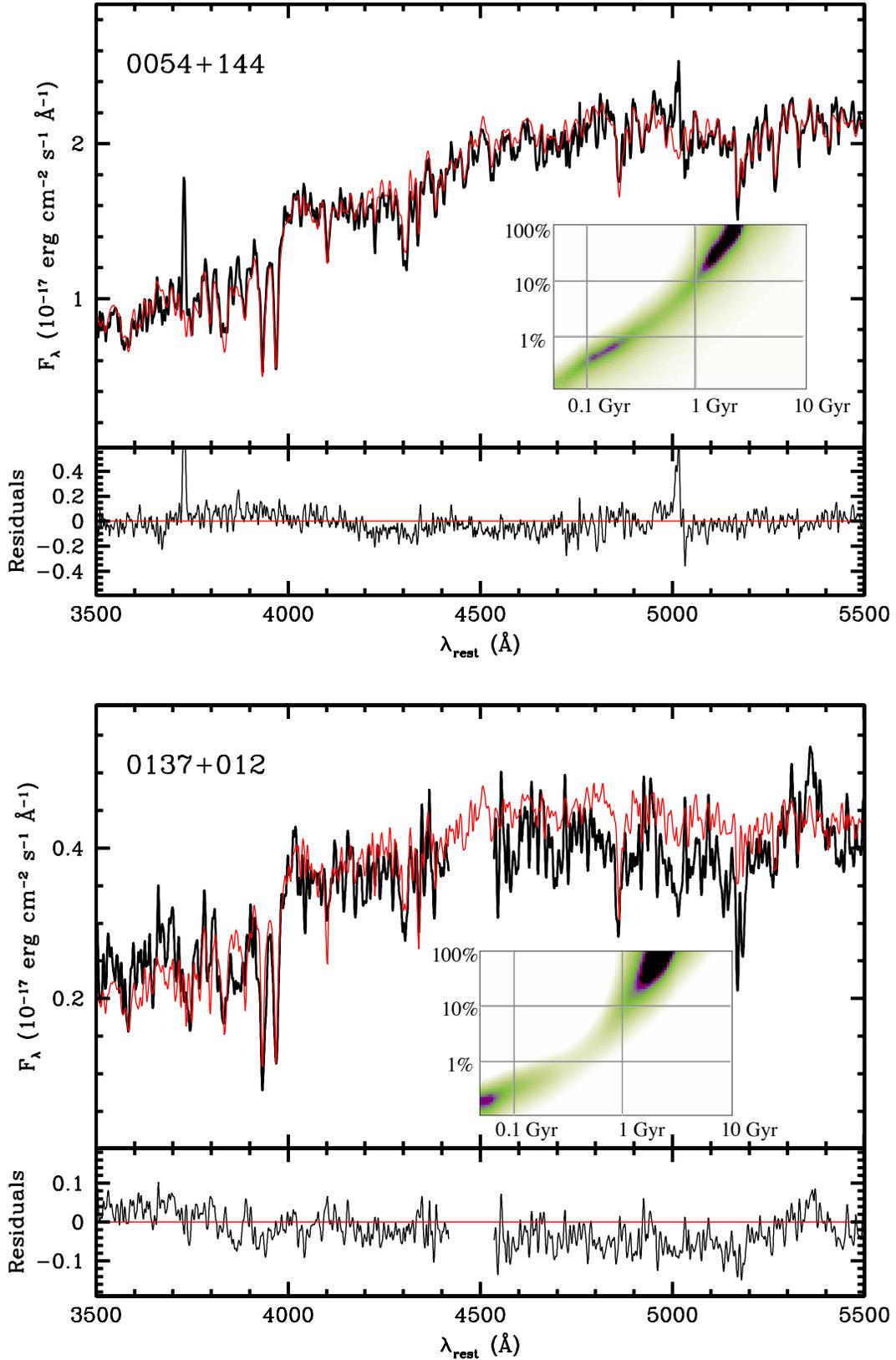}
\caption{Rest-frame Keck LRIS spectra of the QSO host galaxies.  The black trace in the upper panel of each object is the host galaxy corrected for QSO light contamination and for Galactic reddening.   In some cases (0054+144, 0137+012, 0157+001, 1012+008, 2135$-$147, 2247+140), the spectra have been smoothed with a gaussian of sigma $<$1 \AA\ for displaying purposes. The red trace is the best-fit BC03 model to the data, consisting of the combination of a 10 Gyr-old population and a younger population.  The age and relative contribution of this younger population are given in Table~\ref{ages}.  The bottom panel for each object shows the residuals obtained by subtracting the model from the observed spectrum.  The insets are two-dimensional plots of $\chi^{2}$ as a function of the age and relative contribution by mass of the younger population to the total spectrum.}
\label{figure:bestfit}
\end{figure*}
\begin{figure*}[p]
\figurenum{1}
\plotone{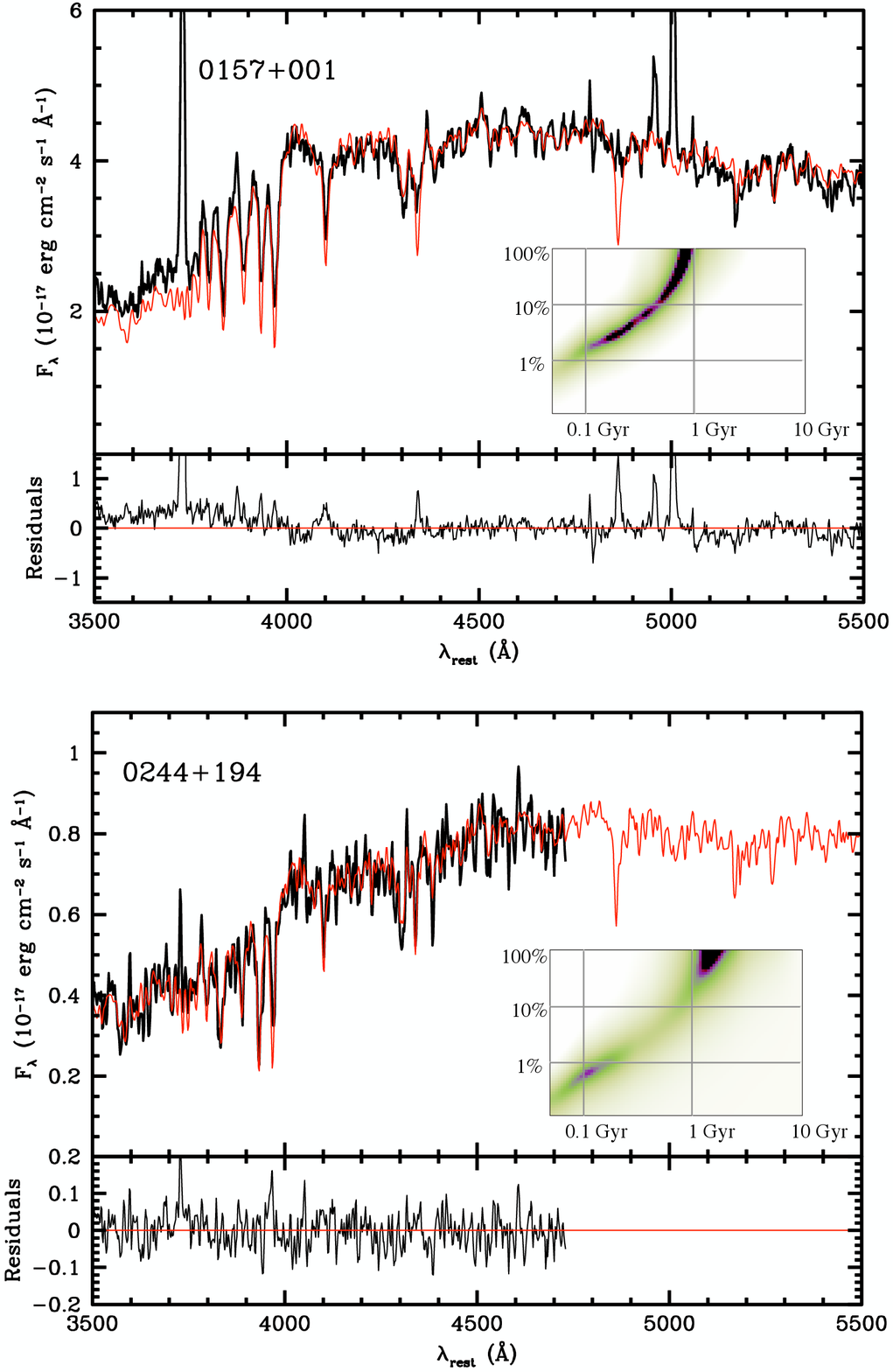}
\caption{Continued.}
\end{figure*}
\begin{figure*}[p]
\figurenum{1}
\plotone{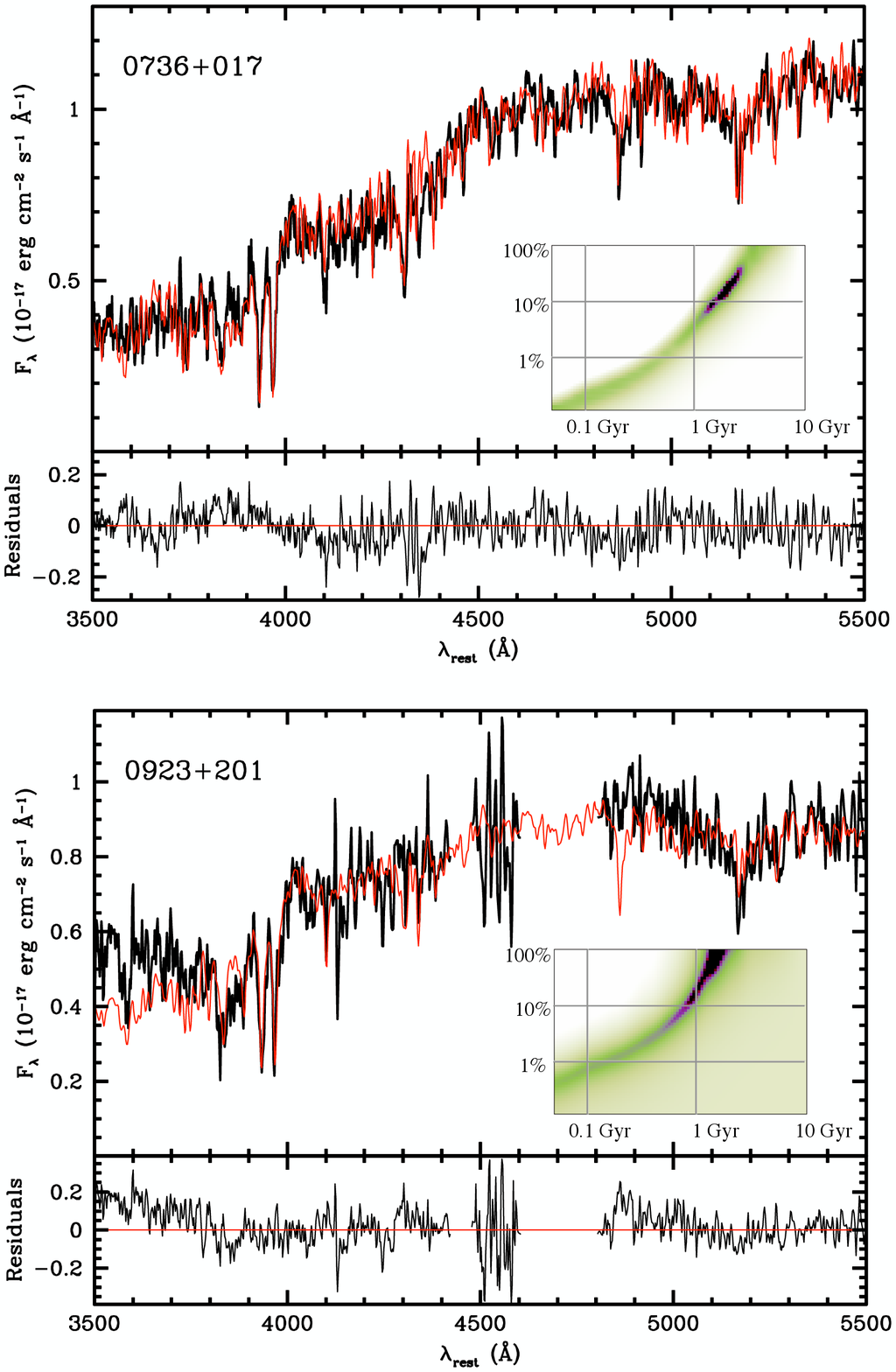}
\caption{Continued.}
\end{figure*}
\begin{figure*}[p]
\figurenum{1}
\plotone{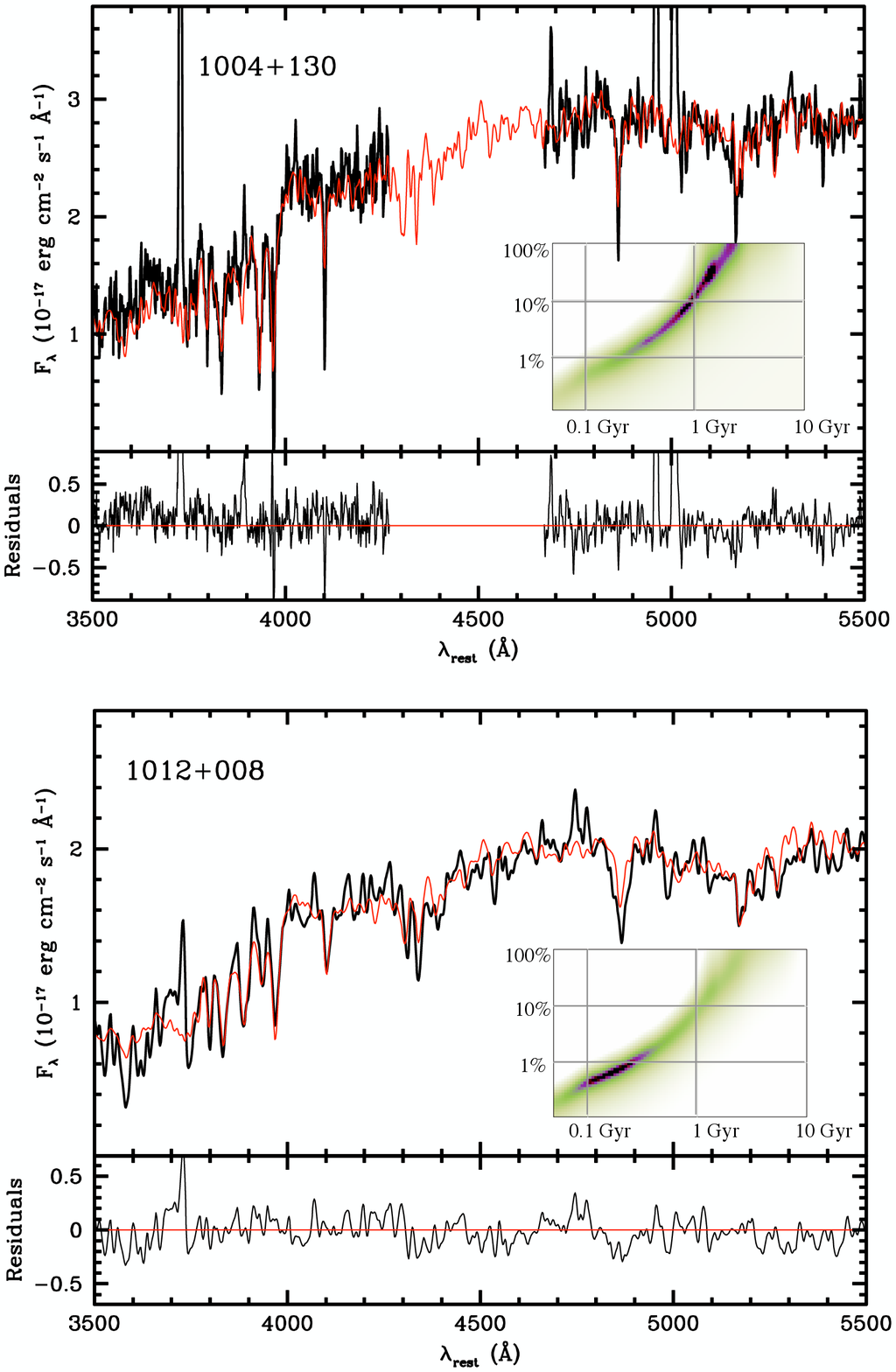}
\caption{Continued.}
\end{figure*}
\begin{figure*}[p]
\figurenum{1}
\plotone{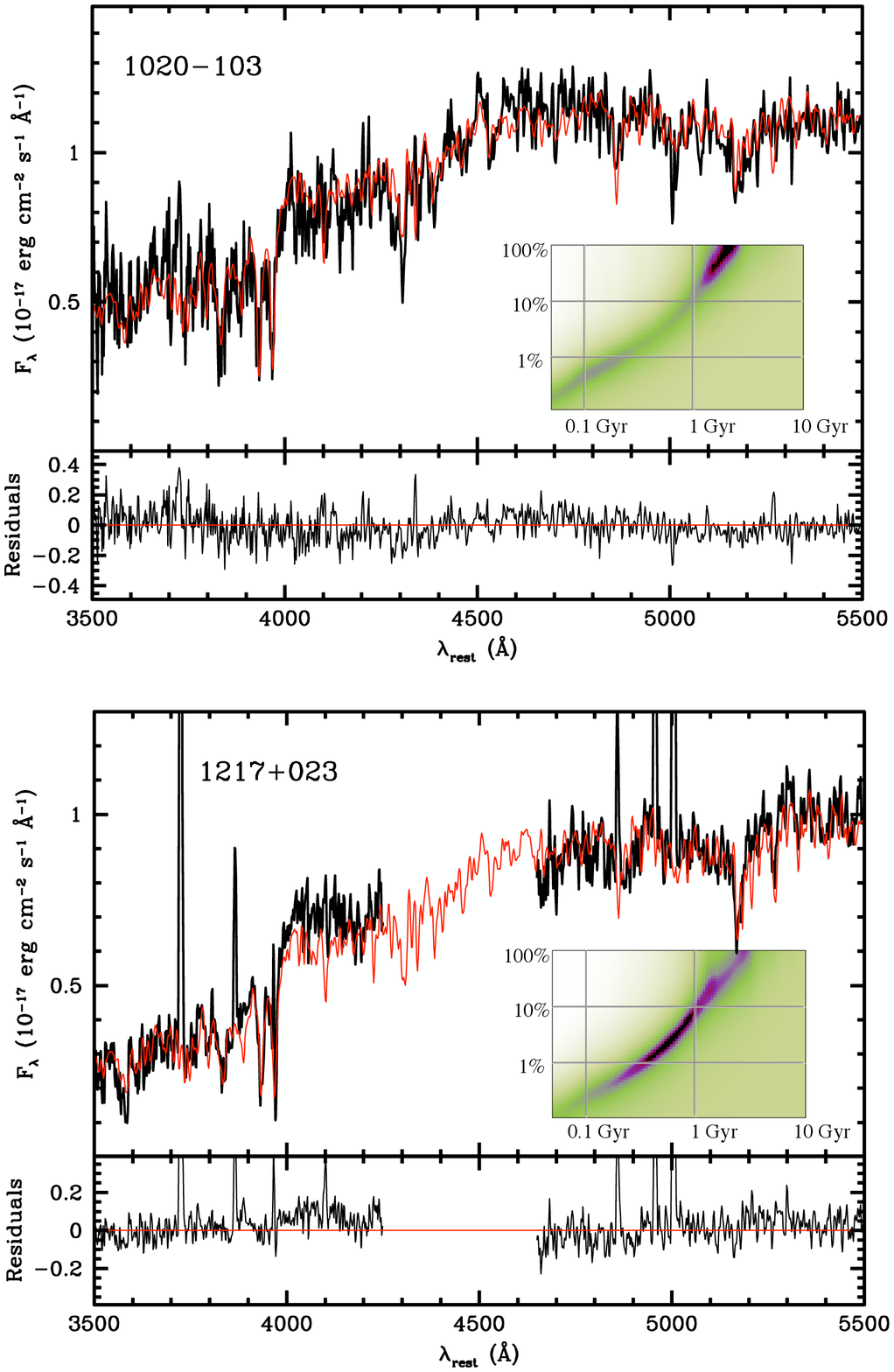}
\caption{Continued.}
\end{figure*}
\begin{figure*}[p]
\figurenum{1}
\plotone{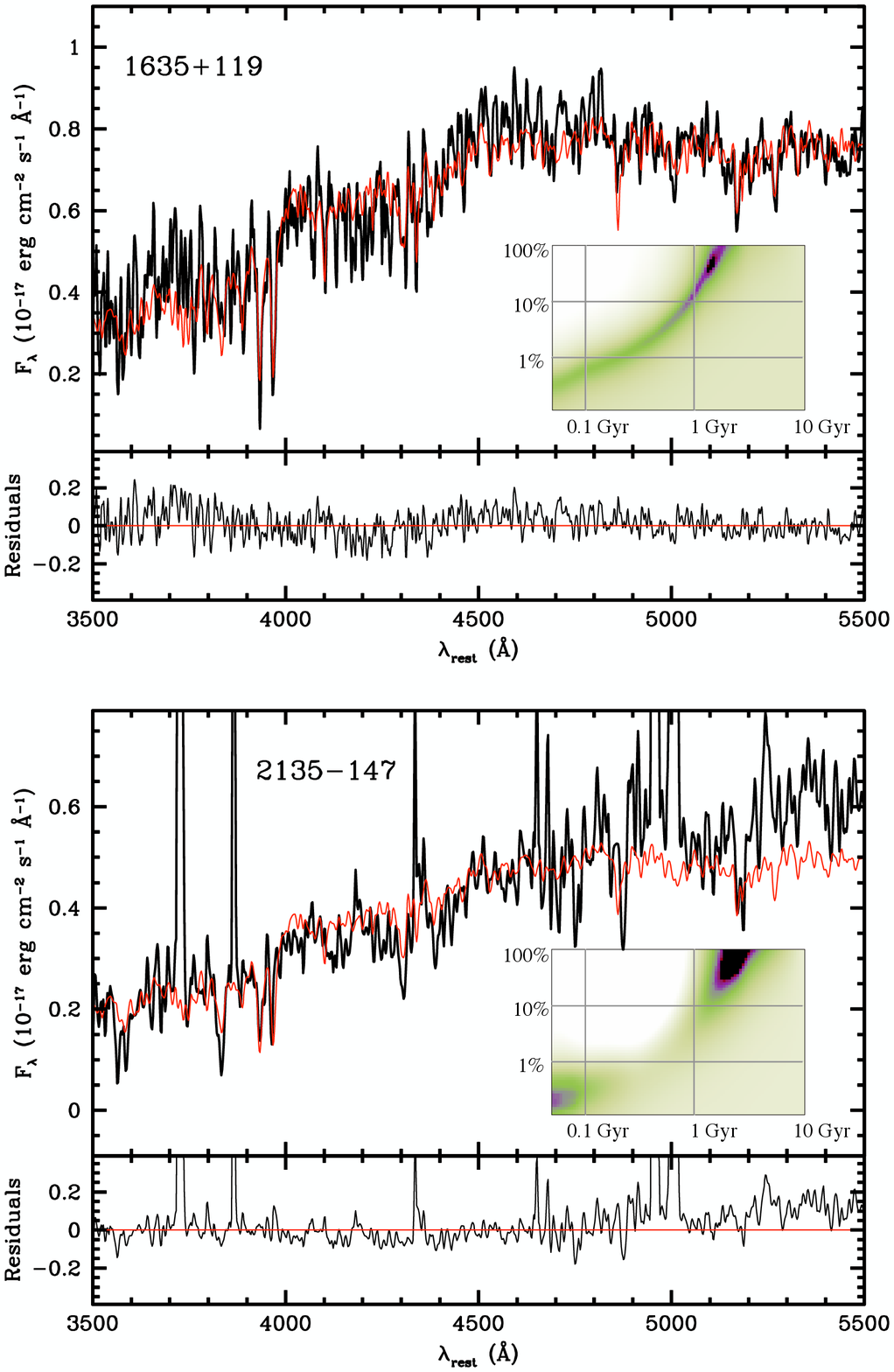}
\caption{Continued.}
\end{figure*}
\begin{figure*}[p]
\figurenum{1}
\plotone{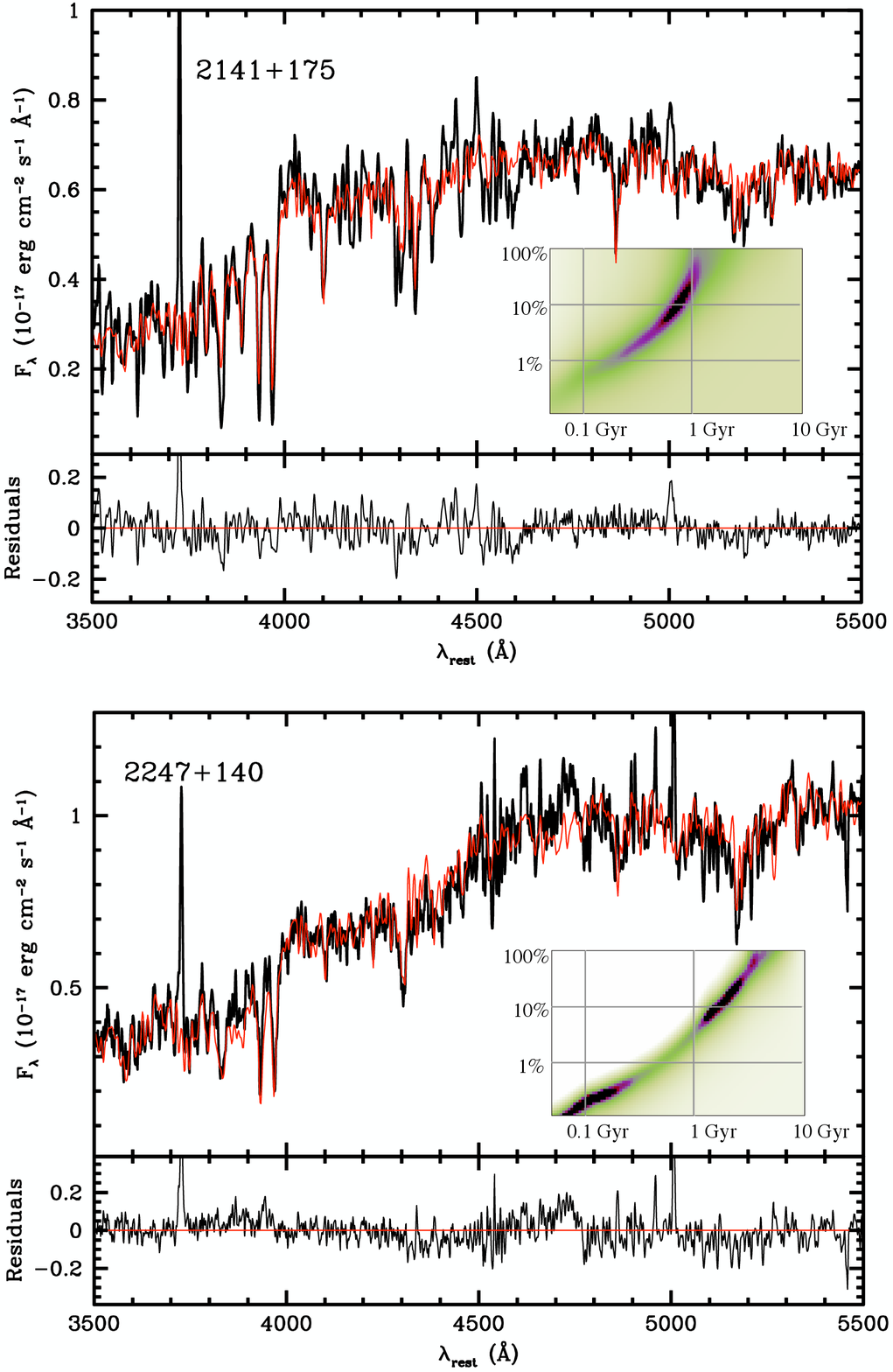}
\caption{Continued.}
\end{figure*}
\begin{figure*}[t!]
\epsscale{1.2}
\figurenum{1}
\plotone{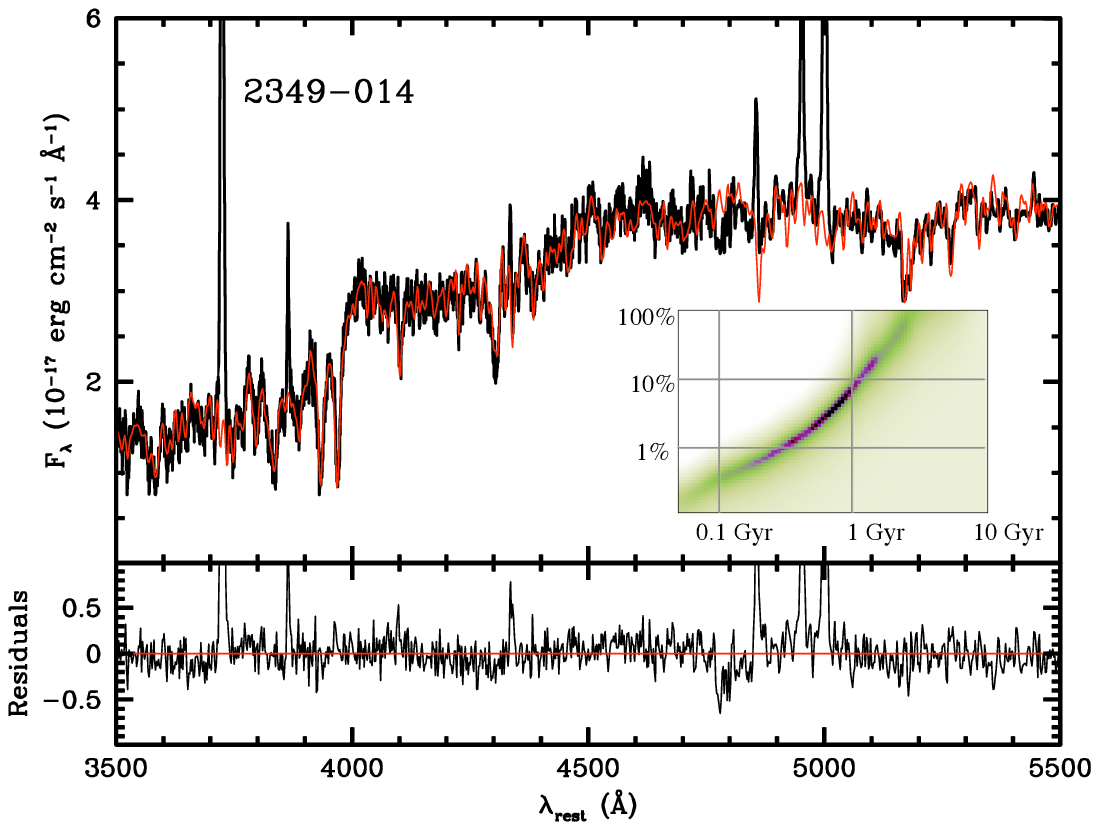}
\caption{Continued.}
\end{figure*}

\begin{deluxetable*}{lcccccc}
\tablewidth{4.6in}
\tablecaption{Stellar Populations \label{ages}}
\tablehead{\colhead{} & \colhead{QSO} & \colhead{$\sigma_{v}$$^{b}$} & \colhead{Starburst Age} & \colhead{Range}
& \colhead{Contribution} & \colhead{Range} \\
\colhead{Object} & \colhead{Cont.$^{a}$} & \colhead{(km s$^{-1}$)}& \colhead{(Gyr)} & \colhead{(Gyr)}
& \colhead{by mass} & \colhead{} }
\startdata
0054+144&\phn9\%& 160.0 $\pm$ \phn9.6 & 2.1  & 1.3--2.7   & 30\%     & 13--70\% \\ 
0137+012 &  39\% & 160.6 $\pm$ 24.5 & 1.7  & 1.4--2.7   & 50\%     & 20--100\%  \\ 
0157+001 & 21\% &169.4 $\pm$ 65.8 &0.8 & 0.2--1.0   & 61\%   & 3--100\%   \\
0244+194 & 29\% & 122.2 $\pm$ 11.2 & 1.4  & 1.0--1.9   &  90\%    & 50--100\%  \\ 
0736+017 & 26\% & 214.2 $\pm$ 19.6 & 2.2  & 1.6--2.4   &  15\%    &  9--24\%  \\
0923+201 & 57\% & 231.4 $\pm$ 50.2 & 1.4  & 0.8--1.9   &  56\%    &  10--100\%  \\
1004+130 & 50\% & \phn99.6 $\pm$ 59.6 & 1.1  & 0.8--1.4   &  23\%    &  7--29\%  \\ 
1012+008 & 40\% & 157.0 $\pm$ 357\phd & 0.2 & 0.10--0.25 &  0.7\%   &  0.5--1.2\%\\ 
1020$-$103&29\%& 318.3 $\pm$ 73.4 & 1.9 & 1.4--2.4   & 80\%     &  30--100\%  \\ 
1217+023 & 55\% & 239.5 $\pm$ 27.7 & 0.7  & 0.4--0.9   &  3.7\%   &  1.0--6.7\% \\ 
1635+119 &\phn8\%& 150.6 $\pm$ 35.2 & 1.5  & 1.2--1.8   & 42\%     & 25--86\%   \\ 
2135$-$147&48\%& 184.7 $\pm$ 55.4 & 2.4 & 1.8--2.8   & 90\%     &  85--100\%  \\ 
2141+175 & 77\% &  189.8 $\pm$ 20.7 & 0.8  & 0.6--1.0   & 12\%    &  6.3--22\%  \\ 
2247+140 & 20\% &  210.5 $\pm$ 15.7 & 2.1 & 1.2--2.6   & 15\%     &  6--30\%    \\ 
2349$-$014&26\% & 228.4  $\pm$ 14.6 & 0.8 & 0.5--1.1   & 4.2\%    &  2--10\%    \\ 
\enddata
\tablecomments{$^{a}$QSO contribution to total flux of observed off-nuclear host galaxy spectrum, as measured at rest wavelength 4500 \AA.  This contribution has been subtracted as explained in Section~\ref{spectroscopy}. $^{b}$Off-axis stellar velocity dispersions have not been aperture corrected.}
\end{deluxetable*}

In Table~\ref{bcmaraston} we compare the results obtained using the BC03 models with the CB07 and M05 models.   In spite of the differences in these sets of models, the best fit starburst ages obtained from these three sets of models are in remarkable agreement.    There is considerably more variation among the models in the values of the relative contribution of the starbursts.  This is not surprising considering the large ranges of values that are given as confidence levels in Table~\ref{ages}.   
The large uncertainties in the relative contributions are explained as follows: While the total flux emitted per solar mass changes rapidly (by more than a factor of two) between a population of 1 and 2 Gyr, the spectral features (including the shape of the continuum) change much more slowly and, therefore, $\chi^{2}$ changes relatively slowly.   In other words, a small increase in age for the starburst requires a larger increase in its relative contribution toward the total spectrum.   
This results in a broad range of values for the contribution that give similar values of $\chi^{2}$ (see insets in Fig.~\ref{figure:bestfit}).  Thus, the precise values of the relative contribution should not be taken at face value, but rather as a general guide for the relative importance of the starburst population.  Even taking into account the large uncertainties, our results show that the contribution from the intermediate-age populations is rather substantial.

\begin{deluxetable*}{lcccccc}
\tablecaption{Comparison of BC03, CB07, and M05 Models \label{bcmaraston}}
\tablehead{\colhead{} & \multicolumn{2}{c}{ Bruzual  \& Charlot '03} & \multicolumn{2}{c}{Charlot \& Bruzual     '07} &\multicolumn{2}{c}{Maraston '05} \\
\colhead{} & \colhead{Starburst Age} & \colhead{Contribution} & \colhead{Starburst Age}
& \colhead{Contribution} & \colhead{Starburst Age} & \colhead{Contribution} \\
\colhead{Object} & \colhead{(Gyr)}
& \colhead{by mass} & \colhead{(Gyr)} & \colhead{by mass} }
\startdata
0054+144 &  2.1 & 30\% & 2.1 & 30\% & 1.5  & 50\%         \\ 
0137+012 & 1.7 & 50\% & 1.7 & 50\% & 1.8  & 65\%         \\ 
0157+001 & 0.8 & 71\% & 0.8 & 77\%&  1.0    & 100\%       \\  
0244+194 & 1.4 & 90\% &1.4 & 90\% & 1.5  & 100\%        \\ 
0736+017 & 2.2 & 15\% & 2.2 & 10\% & 2.0  &  14\%        \\ 
0923+201 & 1.4 & 56\% & 1.0 & 20\% & 0.6  &  34\%        \\  
1004+130 & 1.1 & 23\% &1.1 & 28\% & 0.9  &  20\%        \\  
1012+008 & 0.2 & 0.7\% &0.2 & 0.8\%& 0.6  &  4.0\%       \\  
1020$-$103& 1.9 & 80\% &1.9 & 80\% & 1.0  &  32\%        \\  
1217+023 & 0.7 & 3.7\% &0.5 & 0.8\%& 0.6 &  1.5\%        \\ 
1635+119 & 1.5 & 42\% &1.4 & 55\% & 1.0  & 45\%         \\ 
2135$-$147&2.4 & 90\% & 2.4 & 88\% & 3.0  & 60\%          \\ 
2141+175 & 0.8 & 12\% &0.7 & 9.1\%& 0.6 &  9.3\%         \\ 
2247+140 & 2.1 & 15\% &2.2 & 17\% & 2.0 & 30\%           \\ 
2349$-$014& 0.8 & 4.2\% &1.0 & 7.2\%& 0.8  & 5.5\%         \\ 
\enddata
\end{deluxetable*}

\begin{figure}
\begin{center}
\figurenum{2}
\plotone{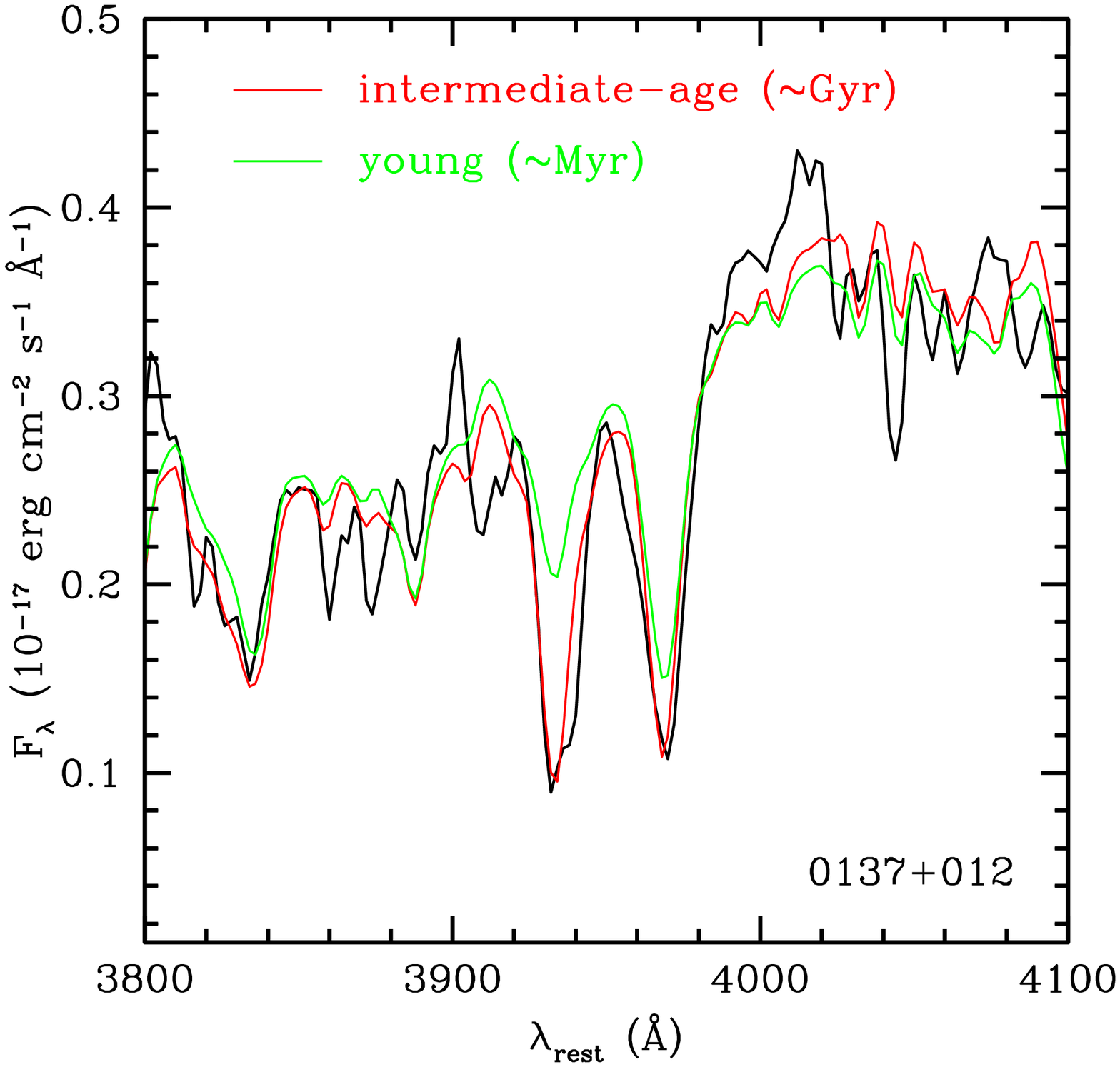}
\caption{Comparison between two models that fit well the continuum shape of a host galaxy.   The black trace is the rest-frame spectrum of the 0137+012 host.   The green trace is the sum of a 10 Gyr population and a 40 Myr population, contributing 0.3\% of the total mass.   The red trace is the sum of a 10 Gyr population and a 1.7 Gyr population contributing 50\% by mass.  Although both models result in a good fit to the continuum, the model including the younger population does not match stellar features such as Ca H\&K.}
\label{figure:hk}
\end{center}
\end{figure}

Six of the objects (0054+144, 0137+012, 0244+194, 1020$-$103, 2135$-$147, 2247+140) have a second minimum in $\chi^{2}$ (see insets in Fig.~\ref{figure:bestfit}).    The second minimum corresponds to a model with a very small contribution from a very young (tens of Myr or younger) population.   This combination of populations reproduces the general shape of the continuum of the host galaxies quite well, and hence the lower value of $\chi^{2}$.  However, close examination of the stellar absorption features reveals that the model corresponding to the primary minimum is, indeed, a much better fit to the data than the values of $\chi^{2}$ would seem to indicate.     Figure~\ref{figure:hk} illustrates this point.   While the continuum of the 0137+012 host is well fit by a model composed of a 40 Myr plus a 10 Gyr old population, the stellar absorption features are best fit by a model including a 1.7 Gyr population.

In our analysis, we did not account for the effects of potential dust in the host galaxies.  Any dust extinction will redden the continua and will therefore lead to overestimating the age and/or metallicity when fit with unreddened models.  Therefore, the ages that we determine for the intermediate-age populations should be regarded as upper limits.   However, the fact that we get similar ages for the intermediate-age populations from techniques that include the overall continuum SED and those that depend primarily on the absorption lines indicates that these populations cannot be substantially younger than these limits.

There was only one object, 2247+140, for which we found a significant age-metallicity degeneracy.  For this object, increasingly older populations with increasingly lower metallicities result in equally good fits.  At the lowest extreme, the spectrum is well fit by a single BC03 model of a 10 Gyr population with a metallicity of 0.4~Z$_{\sun}$.  The M05 models show a similar trend, but in that case the extreme is a fit by a single 1 Gyr population of 0.5~Z$_{\sun}$.  

As noted in the Appendix, 1012+008 presents special challenges, in that the host galaxy spectrum is slightly contaminated by the interacting spiral galaxy, and no observations of a star to calibrate the wavelength variation in scattering of the QSO light were available. Nevertheless, the qualitative presence of an intermediate-age population is quite secure because of the strength of the Balmer absorption lines.

\begin{figure}[t!]
\vspace{5mm}
\epsscale{1.3}
\figurenum{3}
\plotone{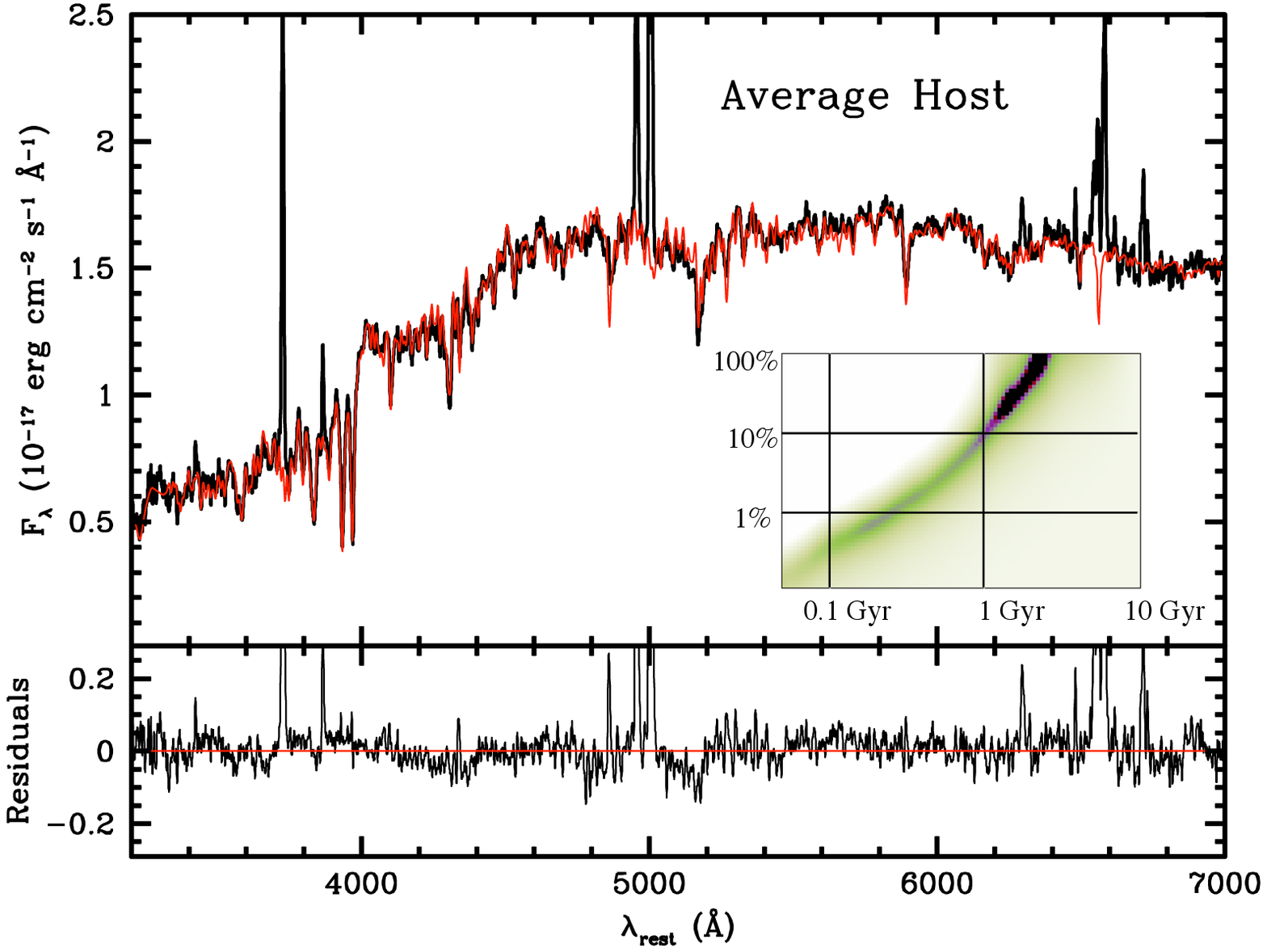}
\caption{Average of the 14 QSO host galaxies with intermediate-age populations.  The red trace is the sum of a 10 Gyr population and a 2.1 Gyr population contributing 64\% by mass along the line of sight.
The inset is a two-dimensional plot of $\chi^{2}$ as a function of the age and relative contribution by mass of the younger population to the total spectrum.}
\label{figure:avghost}
\end{figure}

We created an average QSO host galaxy spectrum (Figure~\ref{figure:avghost}) by combining the spectra of all the host galaxies (except for 1012+008, for the reasons given above).
The spectra were normalized to their median flux in the region between 4020 and 4080 \AA\ and weighted by their S/N.  The resulting average spectrum is best fit by the combination of a 10 Gyr population with a 2.1 (+0.5, $-$0.7) Gyr population contributing 64\% (+36, $-$36) to the total model by mass.   A similar, but inferior, fit may be achieved for the 0.4 Z$_{\sun}$ models with a 10 Gyr plus a 2.0 Gyr population contributing 18\% of the total mass.

Thus, our results indicate that, far from being passively evolving elliptical galaxies, almost all of the QSO host galaxies have suffered a major event of star formation within the past $\sim2$ Gyr.

\subsection{Alternatives}

We have found that the majority (14/15) of the host galaxies in our sample are composed of large fractions of intermediate-age stellar populations.   Here we consider alternative scenarios that may be physically plausible.   

\subsubsection{Purely old populations}\label{oldpop} 
\begin{figure*}
\begin{center}
\figurenum{4}
\plottwo{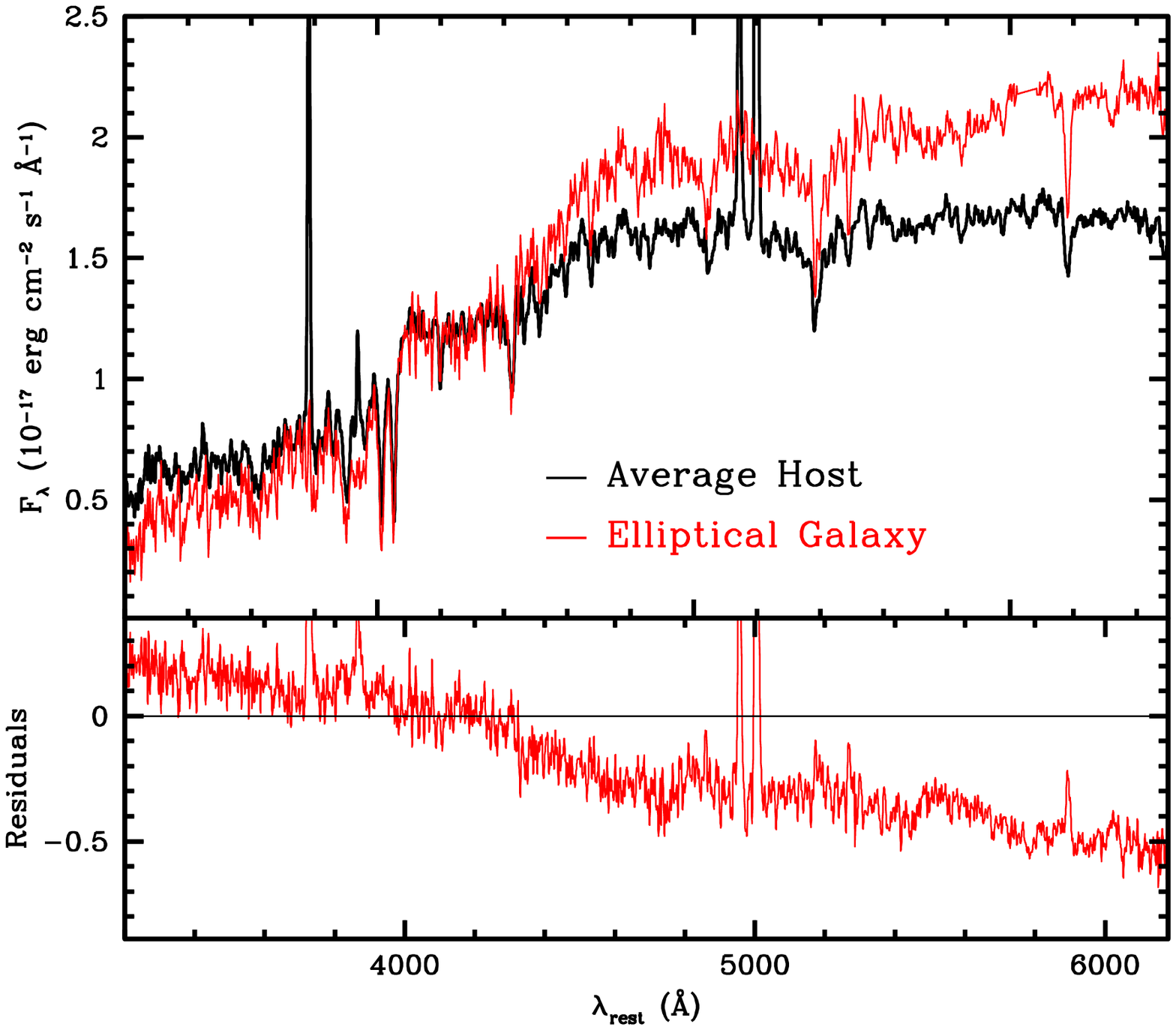}{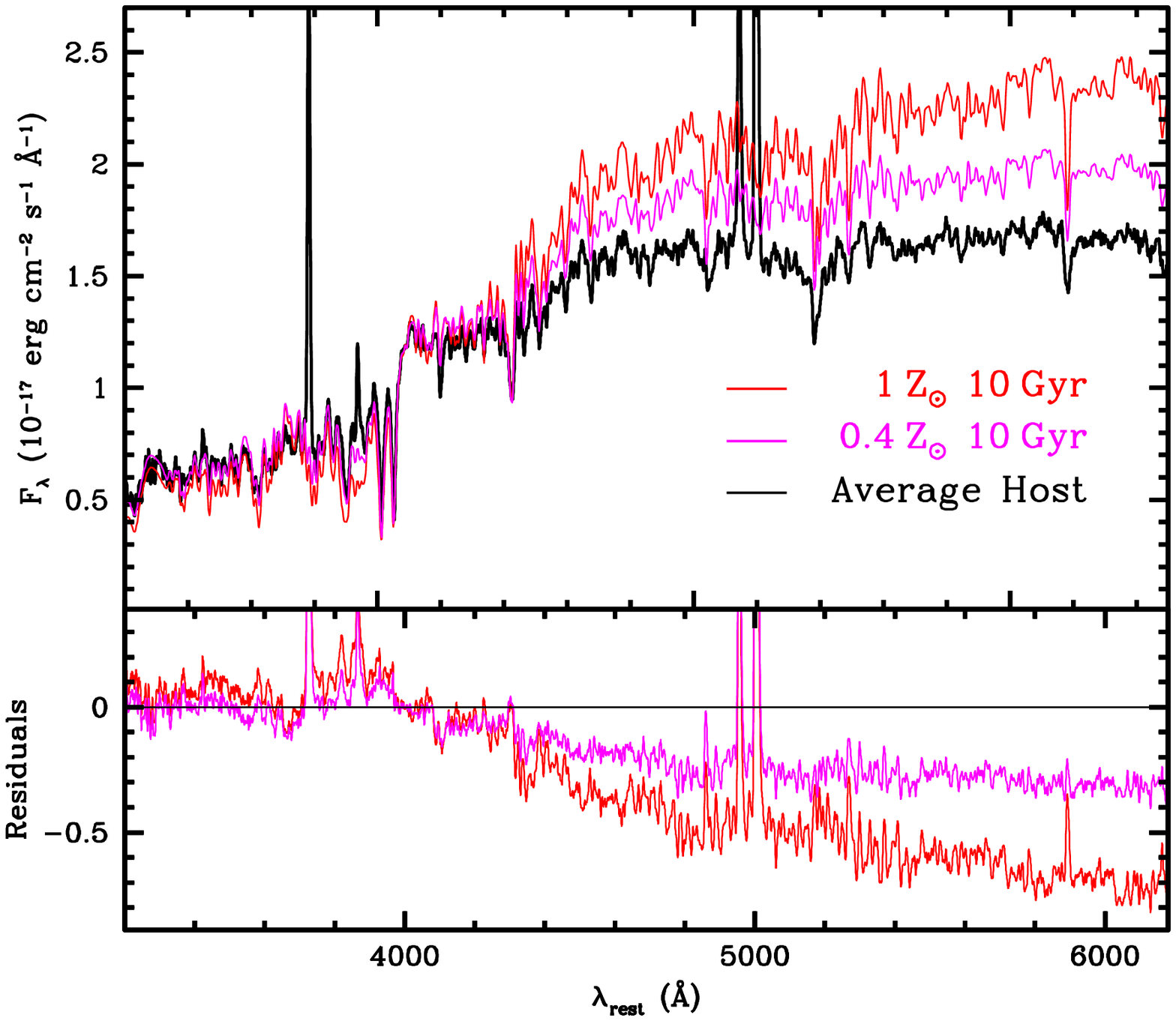}
\vspace{-1cm}
\caption{Comparison between QSO host galaxies and purely old stellar populations.  The black trace is the rest-frame average of 14 QSO host spectra.  The elliptical galaxy spectrum (red trace in the left panel) and the 10 Gyr models (red and purple traces in the right panel) are normalized to the flux of the host galaxy spectrum at 4000 \AA.}
\label{figure:oldpop}
\end{center}
\end{figure*}

Our results clearly rule out the possibility that these hosts are ``red and dead'' elliptical galaxies.  The  $\chi^{2}$ values that we obtain by fitting a single population older than $\sim$8 Gyr are at least four times higher than those obtained from fits including an intermediate-age population.   Figure~\ref{figure:oldpop} shows the average host galaxy spectrum described in Section~\ref{results} above.    For comparison, we over-plot the Keck LRIS spectrum of a z=0.1930 elliptical galaxy from a control sample (G. Canalizo et al., in preparation), normalized to the flux at 4000 \AA.  We also over-plot single 10 Gyr bursts of 0.4 and 1 Z$_{\sun}$.   Neither the elliptical galaxy, nor the different models match the host galaxy spectrum.   The right panel of Figure~\ref{figure:oldpop} shows that the stellar population would have to have a metallicity lower than 0.4 Z$_{\sun}$ in order to match the continuum of the host galaxy spectrum and, even then, the stellar features would not be a good fit to the data, as we discuss below.   In any case, it is highly unlikely that galaxies in this mass range \citep[$\sim10^{11}-10^{12}$ M$_{\odot}$;][]{dunlop2003,decarli2010} would have such low metallicities.

One might argue that the difference between the host galaxy spectra and a purely old population may be explained if the continuum of the host galaxy is affected by QSO light, whether scattered by free electrons and dust, as \citet{young2009} suggest, or simply resulting from poor corrections of the QSO light contamination on the host.     It is clear from the shape of the residuals in Figure~\ref{figure:oldpop} that the difference in the continuum between an old population and our QSO hosts cannot be accounted for by QSO light. 

\begin{figure*}
\begin{center}
\figurenum{5}
\epsscale{1.1}
\plotone{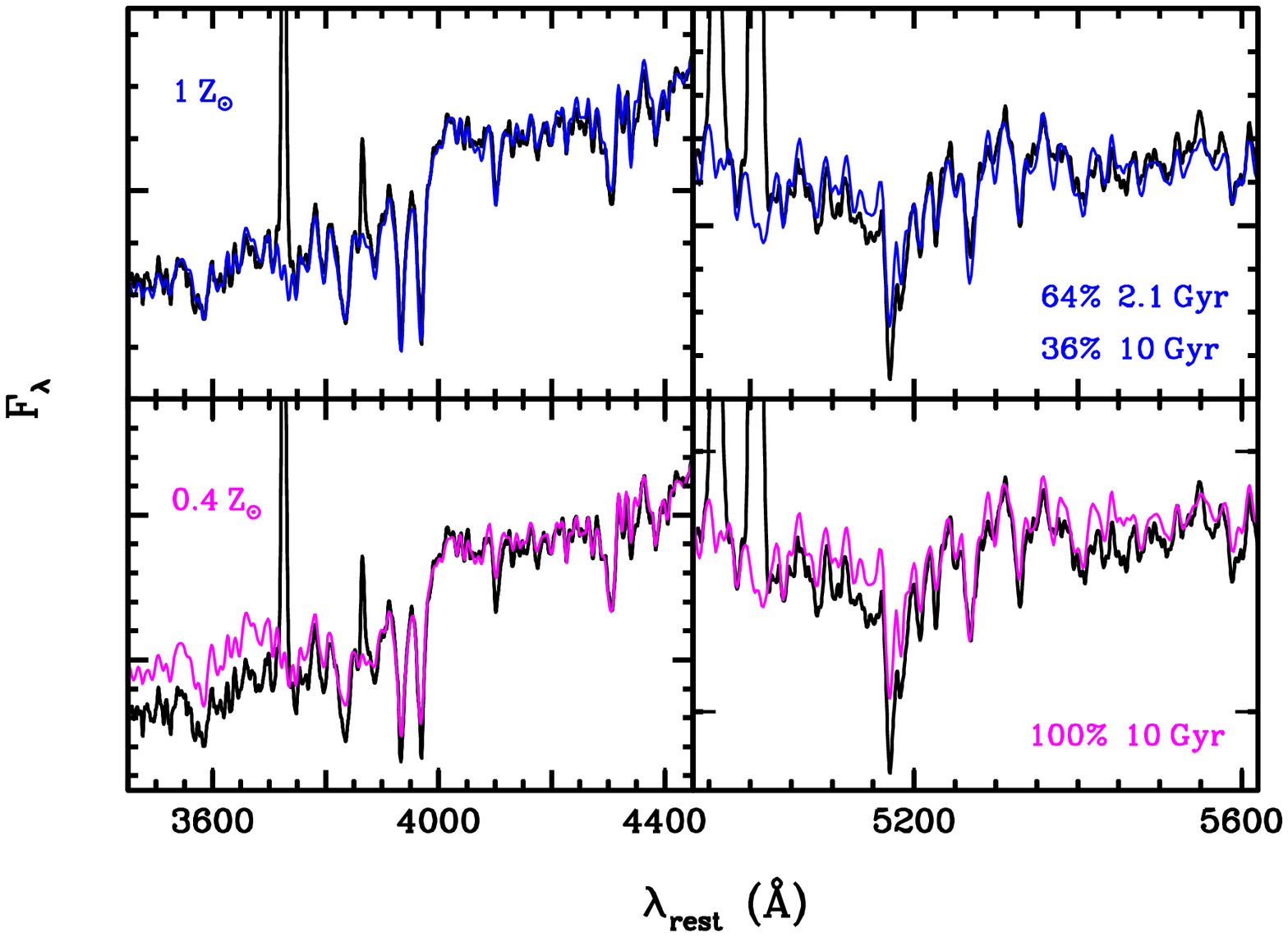}
\caption{Comparison between the average QSO host spectrum and a purely old, low-metalicity  model.   Top two panels: The black trace is the rest-frame average of 14 QSO host spectra.   The blue trace is the best fit 1 Z$_{\sun}$ model for the average host spectrum, composed of a 10 Gyr-old population and a 2.1 Gyr starburst contributing 64\% of the total mass along the line of sight.   Bottom two panels: The black trace is a {\it modified version} of the average host spectrum, created by artificially subtracting a large amount of QSO continuum (with no emission lines) in order to match the SED of the low-metalicity model.  The purple trace is a 10 Gyr-old model with metallicity of 0.4 Z$_{\sun}$.  Refer to Figure~\ref{figure:oldpop} to see the actual difference in SEDs between the average spectrum and the 0.4 Z$_{\sun}$ model.}
\label{figure:continua}
\end{center}
\end{figure*}

To further show that the QSO host spectra differ from old population spectra by more than just the continuum, we subtracted different amounts of QSO continuum from the average host to attempt to match the SED of the purely old populations discussed above.   We subtracted a fit to an average QSO spectrum without including the emission lines, since the latter would otherwise be obviously over-subtracted from the average host.  We found that there is no way to reproduce the SED of either the
elliptical galaxy or the single 10 Gyr-old model of solar metalicity, regardless of the amount of QSO continuum subtracted.   We were able to reproduce roughly the SED from 4000  to 6000 \AA\ of the 0.4 Z$_{\sun}$ 10 Gyr-old model by subtracting an additional 18\% (at 4500 \AA) of QSO continuum (with no emission lines). It is clear from Figure~\ref{figure:continua} that even if the artificially modified average spectrum can match a portion of the SED of a low-metallicity purely old population, the stellar features are better fit by a population that includes an intermediate-age population.
Thus, we conclude that the QSO host galaxy spectra cannot be fit by purely old stellar populations, regardless of metallicity or whether the spectra are contaminated by QSO light.

\subsubsection{A very small contribution from a very recent starburst}
As we mentioned in Section~\ref{results}, the combination of an old population with a very small fraction of a very young (a few Myr- to a few tens of Myr-old) population can mimic the SED of some (6/15) of the hosts in the spectral range that we cover in our observations.    This is essentially the result that \citet{nolan2001} obtained.   However, we have already shown that this particular combination of models is not a good fit to the absorption features in the spectrum.  

Adding any other population to a mostly old (10 Gyr) population results in much worse fits.   We experimented with different possibilities, such as adding a very small amount of sustained star formation rate, or an old population with an exponentially decaying star formation rate with a long e-folding time.  None of these options resulted in acceptable fits.

\subsubsection{Contamination by Blue Horizontal Branch Stars}
\begin{figure}[!tb]
\figurenum{6}
\epsscale{1}
\plotone{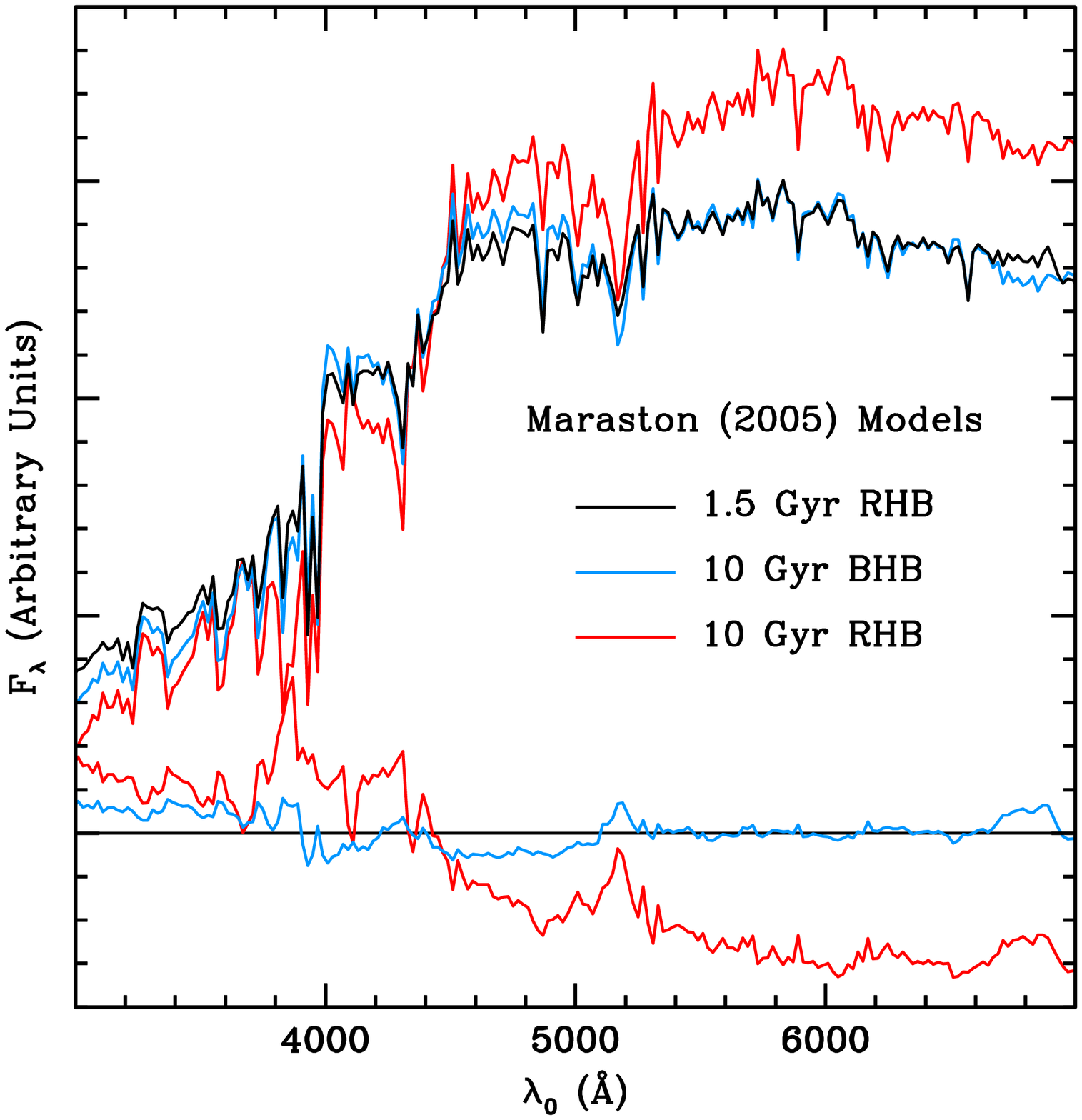}
\caption{Comparison of a 10-Gyr-old stellar population with a blue horizontal branch 
(BHB; blue trace) with a 1.5-Gyr-old stellar population (black trace).  A 10-Gyr-old
stellar population with a red horizontal branch is shown in red for comparison.
All models are from M05 and have solar metallicity.}\label{maraston}
\end{figure}

Can a moderately young (1--2 Gyr) population be impersonated by blue horizontal
branch (BHB) stars in an old population?  Figure \ref{maraston} shows that such a
scenario is at least roughly possible, in that a 10-Gyr-old model with a blue
horizontal branch can mimic a 1.5 Gyr population fairly closely. One now has to
ask whether it is astrophysically reasonable that luminous elliptical galaxies
should have a sufficient population of BHB stars to produce the spectral variations
that we have inferred to be signs of a 1--2-Gyr-old starburst.

The color of a star on the horizontal branch is determined to a large degree
by its metallicity:  the lower the metallicity, the bluer the star. 
By carefully
modeling and subtracting off a maximum old, metal-poor component for
a number of local ellipticals, \citet{trager2005} show that, if a significant intermediate
age population is actually present, correcting for the metal-poor population
does not alter the age obtained for the intermediate-age population by very
much.  More importantly, the presence of the metal-poor population does not
do away with the need for the intermediate-age population to explain the
observed line strengths.

In our spectra, we do not have as good a calibration of line strengths as in
studies of local ellipticals.  On the other hand, for many of our host galaxies,
the Balmer lines are considerably stronger than in the local sample and the inferred intermediate-age
population is much more significant than in the galaxies discussed by \citet{trager2005}. Plausible levels
of BHB contamination are quite insignificant in comparison.   For our few marginal cases, we cannot
completely exclude the possibility that BHB stars might reduce or eliminate the need
for a younger component. However, given the clear presence of intermediate-age populations in much of our sample, it seems more likely that even these are examples of the same trend.

\section{Comparison to other studies}\label{compare}
\citet{nolan2001} use stellar synthesis models by \citet{jimenez2004} to model off-nuclear spectra of the hosts of 14 of the 15 objects in our sample.  Their spectra were obtained with the Mayall 4-m telescope at Kitt Peak and the 4.2-m William Herschel Telescope on La Palma.  In stark contrast to our results, they find that the hosts are dominated by populations of ages 8--14 Gyr, with at most a 1\% contribution from younger stars.   As we mentioned in Section~\ref{method}, they fix a 0.1 Gyr population to account for any recent star formation that may have occurred, and find the best fit by varying the age of the old underlying population.    Given their fitting method and the apparent degeneracy that we found at least for some objects in fitting the shape of the continuum, it is not hard to see how \citet{nolan2001} arrived at this conclusion, especially since their spectra are best constrained redward of the 4000 \AA\ break.   Our spectra have much higher S/N that allow us to discriminate between the two combinations of populations that result into a good fit to the shape of the continuum.

\citet{letawe2007} use Very Large Telescope FORS1 on-axis spectroscopy of the host galaxies of 20 luminous ($M_{B}< -23$) QSOs at $z<0.35$ to study the stellar populations in these objects.  The QSOs are of comparable to slightly higher optical luminosity than the objects in the \citet{dunlop2003} sample (see Figure~1 in \citealt{letawe2007} for a direct comparison of the samples).   The morphologies of the hosts are not exclusively bulge dominated, as in the \citet{dunlop2003} sample, but they also include disks and evidently disturbed hosts.  Using the multi-object mode, they simultaneously observe several PSFs that they then use to deconvolve the 2-D spectra.   They measure Lick indices in the 15 objects for which they are able to recover absorption-line stellar spectra.  By comparing them with the Lick indices of different types of galaxies, they find that only two of their hosts are consistent with being elliptical galaxies, while the rest of them have indices closer to those of late-type galaxies.  They also find that the majority of the hosts harbor large amounts of gas.   

\citet{jahnke2007} conducted a study similar to that of \citet{letawe2007}, but they used a fundamentally different approach to the deconvolution of the host spectra from the PSF.   They perform a semi-analytic two-component spatial PSF fitting of the 2-dimensional  spectra of 18 $z<0.3$ quasars observed on-axis with  EFOSC at the ESO 3.6-m telescope and FORS1 at the ESO VLT.   They then determine stellar population ages of eight of the host galaxies by using two single stellar population BC03 models, including continuous star formation.  They find that the light-weighted age for the hosts is 1-2 Gyr, regardless of whether they are disk or bulge dominated. 

The results of \citet{letawe2007} and \citet{jahnke2007} agree with those of other studies of AGN hosts  based on multi-band imaging studies \citep{kauffmann2003,jahnke2004,sanchez2004}.   While these studies focus on AGN that are, on average, $\sim$1 magnitude fainter than those of the \citet{dunlop2003} sample, they find that the galaxies hosting the most luminous AGN are most often bulge-dominated. However, in agreement with our results and those of \citet{letawe2007}, they find that the host colors are significantly bluer than those of inactive elliptical galaxies, indicating the presence of intermediate-age populations.  In particular, based on their position on the $D_{n}$(4000)/H$\delta_A$ plane, \citet{kauffmann2003} suggest that AGN hosts have had significant bursts of star formation in the past 1-2 Gyr.

\citet{vandenberk2006} use an eigenspectrum decomposition technique to isolate the host galaxy spectra from those of the broad-lined AGN they host in a sample of 4666 objects from the Sloan Digital Sky Survey.  The median redshift of the objects with detected hosts is $z=0.236$.    The majority of the AGN in this sample have lower luminosities than those of the QSOs in the \citet{dunlop2003} sample, but a significant fraction of them have comparable luminosities (see their Figure~13).  \citeauthor{vandenberk2006}  find that the hosts have much higher luminosities than expected for early-type galaxies, and that their colors become increasingly bluer with higher luminosities, when compared with the colors of ellipticals and bulge dominated galaxies.   They define two classification angles formed from the first three galaxy eigencoefficients (equations 2 and 3 in \citealt{vandenberk2006}), which are correlated to spectral type and post-starburst activity, respectively \citep{connolly1995}.    Based on the distribution of the second classification angle in the hosts, they determine that host galaxies have significantly more post-starburst activity than their inactive counterparts.    Although their analysis does not supply specific ages for the stellar populations, their results are at least in qualitative agreement with ours.

\citet{wold2010} obtain off-nuclear Keck LRIS and WIYN SparsePak spectra of 10 QSO hosts at $z<0.3$.   Their sample includes four of the objects in our sample, and six other QSOs with similar optical  luminosities.  Seven of the host galaxies are classified as elliptical.  Their observations are similar to ours and those of \citet{nolan2001}: short exposures of the QSO followed by longer off-axis integrations of the host, with offsets from the nuclei between 2\arcsec and 4\farcs5.  We described their modeling strategy in Section~\ref{method}.     They also find that the elliptical hosts have rest frame $B-V$ colors that are bluer than those of inactive ellipticals.  Moreover, they detect substantial intermediate-age populations in each of the hosts, with an average light-weighted age for the sample of $\sim2$ Gyr.   We are unable to compare directly the precise starburst ages and contributions that we obtain  with the results of \citet{wold2010} because of the way they present their results.   Although they obtain relative contributions of populations with 15 different ages, they bin their results into three groups:  young (age $< $100 Myr), intermediate (age between 100 Myr and 1 Gyr), and old (age $>$1 Gyr).    Since the majority of our objects have starbursts with ages somewhat greater than 1 Gyr, they would be classified as old if we were to use a similar classification.   With this in mind, we find general agreement for the four objects that we have in common with their sample (see Appendix for details).

\citet{cales2013} model the starburst component in the host galaxies of 38 $z\sim$0.3 ``post-starburst quasars'' observed with Keck LRIS and with the KPNO 4-m Mayall telescope and the R-C spectrograph.   These objects show, simultaneously, broad lines from an AGN and absorption lines from a luminous post-starburst stellar population. The sample includes objects with a large range of AGN luminosities, with the high end having significant overlap with our sample.
 \citeauthor{cales2013} use the CB07 models along with models to simulate the AGN contribution.  They do not include a stellar component for a possible old underlying population, as they argue that doing so does not significantly affect the age they determine for the starburst component.   Their results are remarkably similar to ours:  The age of the starburst component in 36/38 of their hosts ranges from 0.5 to 2.6 Gyr, with only two objects having ages of $<$0.3 Gyr.   It is interesting to note that these objects were selected to have intermediate-age populations by design, whereas our objects were selected based on QSO luminosity, yet the two samples of objects have dominant starbursts with the same distribution of ages.

\section{Discussion}\label{discussion}

The studies discussed in the previous section use vastly different techniques to (1) observe the objects, (2) recover the host galaxy spectra, and (3) model the stellar populations.   Yet, with the exception of \citet{nolan2001}, all of these studies point to the presence of a major starburst episode in the host galaxies of low-$z$ QSOs within the last $\sim$2 Gyr.

What caused these massive starbursts and how are they related to the AGN activity?  We have previously reported on the presence of striking tidal features and/or close interacting galaxies for six of the objects in our sample: 0157+001 \citep{canalizo2000b}, 1635+119 \citep{canalizo2007}, 0054+144, 0736+017, 0923+201, 2141+175 \citep{bennert2008}.    The timescales implied by these various features are, in very rough terms, consistent with a picture where mergers were responsible for the starbursts that we observe.  Assuming a projected velocity of 300 km s$^{-1}$ on the plane of the sky for the stars that make up the tail, \citet{canalizo2000b} estimate a dynamical age for the prominent tidal tail in 0157+001 of 330 Myr, compared to the stellar population of 0.8 (+0.2, $-$0.6) Gyr that we find (and already reported in \citealt{canalizo2000b}).  The dynamical age could be underestimated if the projected velocity is lower than 300 km s$^{-1}$.   \citeauthor{canalizo2000b} speculate that the intermediate-age population may be the relic of a starburst ignited at an initial passage of the two interacting galaxies.

\citet{bennert2008} estimate a dynamical age between 250 Myr and 1 Gyr for the tidal tails seen in deep $HST/ACS$ images of 2141+175 and 0054+144 .  While this age estimate is of the same order as the starburst that we observe in 2141+175 (0.8 [+0.2, $-$0.2] Gyr), it is younger than the starburst in 0054+144 (2.1[+0.6,$-$0.8] Gyr).  Thus, the starburst in the latter may have been triggered in a previous passage of the interacting galaxies, as in the case of 0157+001.  Similarly, \citeauthor{bennert2008} estimate a dynamical age of $\sim$1 Gyr for the tidal structure (including ripples, tidal arms, and a faint long tidal tail) surrounding the host of 0736+201, which is of the order of, but younger, than the starburst age of $\sim$2.2 (+0.2,$-$0.6) Gyr that we find in the host.

\citet{canalizo2007} find striking shell structure in the host of 1635+119, along with ripples or tails and other debris extending out to a distance of $\sim$65 kpc.  Their N-body simulations allow the structure to be as old as 1.7 Gyr, which is similar to the 1.6 (+0.4,$-$0.2) Gyr starburst age we determine for the host.

While the host of 0923+201 does not show any obvious tidal features, it lies in a galaxy group of at least six members at similar redshifts \citep[][and references therein]{bennert2008}, with several of the galaxies showing clear signs of interaction.   There is one pair of interacting large elliptical galaxies at a projected distance of $\sim$37 kpc from the QSO host, and a smaller galaxy with tail-like structures $\sim$56 kpc away.   It is likely that these galaxies have interacted with the host galaxy within the past 1.4 Gyrs, triggering the star formation in the host.

We will present deep $HST/WFPC2$ images of the remaining objects in the sample, together with similar images of a matched control sample, in a forthcoming paper.  In summary, we find clear traces of merger and interaction events in almost two thirds of the QSO hosts  and hardly any such traces in the control sample.

A few of the spectroscopic surveys discussed in Section~\ref{compare} also examine the morphologies of the hosts.  \citet{jahnke2007} and \citet{letawe2007} find that $\sim$50\% of the hosts in their samples show morphological signs of interactions and disturbed radial velocity curves.  Of the 29 post-starbursts quasars that have $HST$ images \citep[presented by][]{cales2011}, 17 ($\sim$59\%) show signs of interactions or mergers.  The early-type galaxies in the sample, which host the most luminous AGN, have a similar fraction of objects with signs of interaction.  

Although at first it would seem that the fractions of interacting objects in these samples are lower than in ours, we must keep in mind that our results are derived from high-resolution, very deep (five orbits) $HST$ images that allow us to detect the aging tell-tale signs of mergers as they are fading away.   If we were to analyze shallow and/or ground-based images of the six objects in our sample discussed above, we would find that only two (0157+001 and 2141+175) have obvious signs of tidal interactions and one (0923+201) has close interacting galaxies.   Thus, based on this subsample, we would conclude that only 50\% of the hosts are connected to mergers.   It is then possible that most, if not all, of the QSO host galaxies that appear to be bulge dominated may indeed be more evolved merger remnants hosting intermediate age populations.

Thus, we find compelling evidence that the aging starbursts we observe in early-type QSO hosts were triggered during the early stages of a merger event.   Typical QSO lifetimes are expected to be, at most, $\sim10^8$ years, so the currently observed QSO activity cannot have been triggered at the same time as these $\sim10^9$-year-old starbursts. How, then, can we explain this correlation between QSO activity and intermediate-age starbursts?

Numerical simulations of merging galaxies show that the amplitude and duration of enhanced star forming activity, as well as the stage at which it occurs during a merger, depend on the specific parameters of the encounter (orbits, mass ratios, etc.) as well as the morphological type of the progenitors \citep[e.g.,][]{dimatteo2007}.   It is clear, however, that many mergers result in two major epochs of star formation: one soon after first passage, and one during final coalescence \citep[e.g.,][]{mihos1996,cox2008,torrey2012,vanwassenhove2012}.   Similarly, numerical simulations show that accretion onto the central black hole(s) is episodic, and there can be two (or more) major epochs of AGN activity roughly corresponding to (but somewhat delayed with respect to; see \citealt{hopkins2012}) the times at which star formation is dramatically enhanced \citep[e.g.,][N.R.\ Stickley and G. Canalizo, in preparation]{hopkins2008,vanwassenhove2012}.   The time delay between the first and second epochs of enhanced star formation/AGN activity naturally depends on the details of the merger; typical values are of the order of 1 Gyr, i.e., of the same order as the ages of the starbursts that we observe in the QSO hosts.

Thus, our observations appear to be consistent with the predictions made by numerical simulations.  In this picture, the QSO hosts in our sample are products of galaxy mergers.   Strong bursts of star formation are induced in the early stages of a galaxy interaction.  As the galaxies continue to merge for the next billion or so years, the new stars mix with the older populations, and they age as the morphology of the newly formed galaxy begins to relax.   When gas continues to concentrate in the central regions of the galaxy, a new central starburst is triggered, followed by a final epoch of AGN activity.   It is precisely at this moment that we seem to be observing the QSOs in our sample.

A glaring difference between this picture and our results is the absence of young stellar populations, expected to be the result of the major starburst triggered during the final coalescence of the galaxies, and a short time (of the order of $\sim$100 Myr; \citealt{hopkins2012}) before the triggering of the QSO.   However, these starbursts are expected to occur very close to the nucleus (P. Hopkins, S. Van Wassenhove, private communication), whereas our spectroscopic observations cover regions of the host galaxies that are typically $\sim$8 kpc away from the center (see Table 1). 
Stars formed during earlier stages of mergers are more easily mixed with the rest of the population by the end of the merger because of the multiple passes that the interacting galaxies experience.   However, once the galaxies merge, there is no further violent event to help mix the new stars.   This is supported by numerical simulations that show that the velocity dispersions of stars formed in the late stages of mergers remain low compared to those of stars formed earlier on (N.R.\ Stickley and G.\ Canalizo, in preparation).

At least for the one object for which we have spatially resolved spectra, 0157+001 \citep{canalizo2000b}, we do find significantly younger populations ($\sim$200 Myr old) near the QSO nucleus.   There are a few other QSO hosts for which spatially resolved information on stellar populations exists.   \citet{canalizo2000a} show that the most luminous and youngest ($<$10 Myr) stellar populations in the host galaxy of 3C\,48 are concentrated in the central regions of the host (in the central $\sim$5 kpc), while the outer regions are dominated by an older population.   Unfortunately, the S/N of the spectra in the outer regions is not sufficient to discriminate between a purely old population and a population including intermediate-age stars.    Similarly, the youngest starbursts ($<$50 Myr) in the host of Mrk\,231 are found in a region $\sim$3 kpc from the QSO nucleus \citep{canalizo2000b}, whereas the outer regions are dominated by older populations.   As in the case of 3C\,48, the S/N is insufficient to determine the precise composition of the older populations.   Finally, \citet{brotherton2002} report on the presence of a $\lesssim$40 Myr starburst in the central 600 pc of the host galaxy of the prototype post-starburst quasar, UN\,J1025$-$0040, which is significantly younger than the dominant population of the host of 400 Myr.

However, there is evidence that the picture may be more complicated. \cite{shi2007} found strong evidence that QSOs selected by different criteria showed different levels of star formation. In particular, their sample of PG QSOs \citep[generally radio quiet;][]{schmidt1983} showed significantly more star formation than did their sample of 3CR (i.e., radio-loud) QSOs. This result is amplified by \citet{fu2009}, who found that, of a sample of 13 radio-loud QSOs, only the compact-steep-spectrum quasar 3C\,48 showed any evidence of star formation. For the remaining 12 QSOs, all with FR\,II radio morphologies, a stack of deep {\it Spitzer} IRS spectra indicated a conservative average star-formation rate upper limit of $\sim$12 M$_{\odot}$ yr$^{-1}$. These differences in typical star-formation rates between radio-quiet QSOs and FR\,II quasars most likely reflect differences in environments \citep[][and references therein]{fu2009}, but it is tempting to speculate that there may also be a dependance on merger stage. In any case, when \citet{shi2009} say, ``We propose that type-1 quasars reside in a distinct galaxy population that shows elliptical morphology but that harbors a significant fraction of intermediate-age stars and is experiencing intense circumnuclear star formation,'' the last part of this statement appears not to hold generally for FR\,II quasar hosts.

We have concentrated on the QSO sample of \citet{dunlop2003} precisely because of their contention that the host galaxies of this sample have properties indistinguishable from those of local quiescent ellipticals. We find instead that virtually all of them have intermediate-age populations comprising a significant fraction of the total mass of the host galaxies. 
These populations can be interpreted in a natural way in terms of models of galaxy mergers, where the star formation is triggered by a previous passage of the galaxies, while the current QSO activity results from the final merger. While there certainly are some cases of QSOs seen in conjunction with contemporaneous starbursts, or at least with much younger populations \citep{canalizo2001}, these seem to be relatively rare, and they probably depend on having serendipitous sightlines to the QSO through otherwise dusty host galaxies. In most objects at similar stages, the QSO nucleus is likely obscured, and the galaxies would be classified as ULIRGs or HyLIRGS. By the time the starburst has subsided and the dust has cleared, the QSO activity likely will have stopped, only to be re-triggered during the final merger.

\acknowledgments

We are grateful to the anonymous referee for useful comments and suggestions that helped improve both the contents and the presentation of this paper.
This work was supported in part by the 
National Science Foundation, under grants number AST 0507450 and AST 0807900.  
Additional support was provided by NASA (programs GO-10421, GO-11101, and AR-10941) through a grant from the Space Telescope Science Institute, which is operated by the Association of Universities for Research in Astronomy, Incorporated, under NASA contract NAS5-26555.
G.C.\ is grateful for the hospitality of the Institute for Astronomy, University of Hawaii, where part of this work was conducted. 
The data presented herein were obtained at the W.M. Keck Observatory, 
which is operated as a scientific partnership among the California Institute 
of Technology, the University of California and the National Aeronautics and 
Space Administration. The Observatory was made possible by the generous 
financial support of the W.M. Keck Foundation.   The authors recognize the 
very significant 
cultural role that the summit of Mauna Kea has within the indigenous
Hawaiian community and are grateful to have had the opportunity to
conduct observations from it.  This research made use of the NASA/IPAC 
Extragalactic Database (NED) 
which is operated by the Jet Propulsion Laboratory, California Institute of 
Technology, under contract with the National Aeronautics and Space 
Administration.

{\it Facilities:} \facility{Keck:I (LRIS)}

\appendix{Notes on Individual Objects}

{\bf 0054+144:} (PHL 909, RQQ) The spectrum of this host shows weak  [\ion{O}{3}] $\lambda$5007 and somewhat stronger [\ion{O}{2}] $\lambda$3727 (EW$\sim$7 \AA).  For the M05 models, a better fit is achieved using 0.5 Z$_{\sun}$ models.  In this case, the best-fit younger population is 2.0 Gyr, contributing 59\% of the mass along the line of sight.  \citet{wold2010} find a flux-weighted luminosity of 2.8 Gyr for this object, and a contribution of 43.8\% of the flux by a population of age between 100 Myr and 1 Gyr.   As mentioned in Section~\ref{results}, 0054+144 is one of six objects that have a secondary minimum in $\chi^{2}$ corresponding to a model with a small contribution from a very young population.  In this case, the young population would be 140 Myr, which would be in agreement with the results by \citeauthor{wold2010}, although the flux contribution of this population would be significantly smaller ($\sim$20\%) than that reported by \citeauthor{wold2010} The lower S/N of their spectrum would have prevented them from using absorption features to discriminate between the two models that fit the continuum of this object.  The morphology of this object, as determined from deep $HST$ observations, is described in detail by \citet{bennert2008}.

{\bf 0137+012:} (PHL 1093, RLQ)  Images of this QSO \citep[e.g.,][]{dunlop2003} show a bright knot $\sim$1\arcsec ($\sim$4 kpc) west of the nucleus.  We obtained a 300 s spectrum of this object, which shows that it is a compact galaxy with a similar stellar population to that of 0137+012 and a redshift of $z$=0.2625.  The relative velocity between this companion and the host is 356 km s$^{-1}$.  As also reported by \citet{hughes2000}, there is no evidence for emission lines.

{\bf 0157+001:} (Mrk 1014, RQQ) We described in detail the spatially resolved stellar populations in the host galaxy of this object as well as its morphological features  in \citet{canalizo2000b}.  We created the spectrum that we present in this paper by coadding the spectra covered by regions {\it d},  {\it f}, and  {\it g} shown in Figure~4 of \citet{canalizo2000b}.   The spectral modeling of this object also has a secondary minimum corresponding to a smaller contribution of a younger population.  However, as described in Section~\ref{results} and in more detail in \citet{canalizo2000b}, the model including the younger population shows ``significant discrepancies in the region around the 4000 \AA\ break and the \ion{Ca}{2} K line.''

{\bf 0204+292:} (3C 059, RQQ)  This object was observed at high airmass and we were not able to obtain a reliable QSO-subtracted spectrum.   However, even the poorly subtracted spectrum shows clearly stellar absorption features similar to those of the rest of the objects in the sample.  By normalizing the continuum and fitting for the stellar absorption lines only, we find that the best fit is achieved by a single 1.4 (+0.6, $-$0.2) Gyr population.    Nevertheless, the fit is not reliable and we do not include this object in the analysis.

{\bf 0244+194:} (QSO B0244+194, RQQ)  The red-side spectrum of this object was corrupted.  The blue-side spectrum shows [\ion{Ne}{3}] and [\ion{O}{2}] $\lambda$3727, likely ionized by the QSO.

{\bf 0736+017:} (PKS 0736+017, RLQ) \citet{wold2010} find a flux-weighted age for this object of 3.6 Gyr, with all the flux falling in the their ``old'' ($>$1 Gyr) bin, which is in general agreement with our results.   They find no evidence for emission lines.   Although this is one of the host galaxies with the least amount of emission in our sample, we do see faint [\ion{O}{2}] $\lambda$3727 with an equivalent width EW$\lesssim$3.5 \AA.   This may be indicative of a small amount of current star formation or a low velocity shock.  Deep $HST$ observations of this object are described in detail by \citet{bennert2008}.  Our spectra show that two of the small galaxies surrounding 0736+017 (galaxies ``a'' and ``b'' in Figure~4 of \citealt{bennert2008}) are background projected galaxies.

{\bf 0923+201:} (PG 0923+201, RQQ) We find no evidence of emission lines in the host galaxy of this object.  {\it Spitzer Space Telescope} observations of this object show that it has neither FIR emission nor prominent PAH features \citep{shi2007}.  Deep $HST$ observations of this object are described in detail by \citet{bennert2008}. 

{\bf 1004+130:} (PG 1004+130, PKS 1004+130, RLQ)  This object shows very strong contamination from narrow emission lines from the extended narrow line region (NLR).  In the spectrum shown if Figure~\ref{figure:bestfit}, we have subtracted a model spectrum of the Balmer emission lines in the NLR.  Since this introduces an uncertainty in the fitting of the Balmer absorption lines, we excluded the latter from the fits.  Figure~\ref{figure:bestfit} shows that the NLR model we subtracted is reasonable, although it appears to have somewhat overestimated the depth of the higher order Balmer lines.   The 2-dimensional spectra show that the position of the strong emission is slightly offset from that of the stellar continuum to the SE, supporting the idea that emission is not associated with star formation.  

{\bf 1012+008:} (PG 1012+008, RQQ)  Although the slit was placed through the QSO nucleus, the host galaxy spectrum was extracted from a region $\sim$2$\arcsec$ ($\sim$6.2 kpc) to the SE of the nucleus.  This region corresponds to the bridge between the host and an interacting spiral galaxy seen in $HST$ images \citep[e.g.][]{mclure1999,bahcall1997}, so the spectrum is likely to be contaminated by stars from the interacting galaxy.    This object was observed during a different campaign where we did not obtain separate off-axis observations of PSFs to model the QSO contamination.   As a result, the QSO subtraction is somewhat less reliable for this object than for the rest of the objects, as can be seen from the profile of the H$\beta$ line (Figure~\ref{figure:bestfit}), which is clearly over-subtracted.   Nevertheless, the deep higher-order Balmer lines and the ratio of the \ion{Ca}{2} H\&K lines indicate that the stellar population contains a fairly young (a few hundred Myr old or younger) component. 

{\bf 1020$-$103:} (PKS 1020$-$103, RLQ)  The spectrum of this object shows very weak [\ion{O}{2}] $\lambda$3727 emission, with EW$\lesssim$5 \AA.  $HST$ images show a disturbed spiral  galaxy at a projected distance of $\sim$62 kpc NW ($\alpha_{J2000}$ = 10:22:32.3, $\delta_{J2000}$ = $-$10:37:24) of the QSO with a redshift of z=0.1959, as measured from our spectra.  The relative velocity between the two galaxies is 175 km s$^{-1}$, so they are likely to be gravitationally bound.  

{\bf 1217+023:}  (PKS 1217+02, RLQ)   The spectrum of this host shows strong emission lines, likely from the extended NLR.  There is clear emission contaminating the Balmer absorption lines, so the latter were excluded from the fits.   

{\bf 1549+203:} (1E1549+203, RQQ)  The S/N of the spectrum of this object was too low to allow us to recover the host galaxy spectrum.   The redshift that we report in Table~1 was measured from [\ion{O}{3}] narrow emission lines.

{\bf 1635+119:} (MC2 1635+119, RQQ)  The spectrum of this host galaxy was extracted from a region 2\arcsec$-$5\arcsec (5.1$-$12.8 kpc) from the nucleus on either side of the QSO host.  The spectrum does not show any significant emission lines.   \citet{canalizo2007} describe the morphology of this object.  The age of the intermediate-age population reported by \citeauthor{canalizo2007} is 1.4 (+0.5, $-$0.2) Gyr contributing 52\% to the total population.   This age was estimated using the CB07 models, with a 12 Gyr-old population as an underlying old population.   

{\bf 2135$-$147:} (PKS 2135$-$14, RLQ)  This object has a strong extended NLR \citep[e.g.,][]{stockton1987,canalizo1997}, clearly seen in the spectrum of the host. In the spectrum shown in Figure~\ref{figure:bestfit}, we have subtracted a model spectrum of the NLR Balmer emission lines.  Since this introduces an uncertainty in the fitting of the Balmer absorption lines, we excluded the latter from the fits. \citet{wold2010} report a stellar population for the object $>$ 1 Gyr, with no contribution from younger populations.   They give a light-weighted age of $\sim$3.3 Gyr.   This is roughly consistent with our results. 

{\bf 2141+175:} (PG 2141+175, OX 169, RLQ)  This object shows weak [\ion{O}{3}] $\lambda$5007 and  stronger [\ion{O}{2}] $\lambda$3727 with EW$\sim$23 \AA. This may be indicative of some current star formation in the host.   Deep $HST$ observations of this object are described in detail by \citet{bennert2008}. 

{\bf 2247+140:} (PG 2247+140, PKS 2247+140, RLQ) As discussed in Section~\ref{results}, the modeling for this object presents a significant age-metallicity degeneracy: Increasingly older populations with increasingly lower metallicities result in equally good fits.   The spectrum shows weak 
[\ion{O}{3}] $\lambda$5007 and somewhat stronger [\ion{O}{2}] $\lambda$3727 (EW$\sim$11 \AA).   This could be indicative of some current star formation in the host.

{\bf 2349$-$014:} (PG 2349$-$014, PKS 2349$-$01, RLQ)   For this object, the fit achieved using the 0.4 Z$_{\sun}$ models is slightly better than that obtained using solar metallicity models.   In that case, the age of the starburst is also 0.8 Gyr, but contributes only 2\% of the total mass.  However, for the M05 models, solar metallicity models clearly give the best fit.  \citet{wold2010} model the spectra from two different regions of this host.  Our slit was placed roughly perpendicular to theirs, so we sample a different region, although there is some overlap.   For the spectrum corresponding to the slit closer to the nucleus (3\arcsec S), they find a contribution of 26.4\% by flux from populations $\leq$ 1 Gyr (15.3\% from populations between 100 Myr and 1 Gyr, i.e., their ``intermediate-age'' bin). For the other slit (4\arcsec N), they find a smaller contribution of 7.1\% by flux for these populations (0.8\% from populations between 100 Myr and 1 Gyr).    The 0.8 Gyr starburst that we find contributes $\sim$45\% of the total flux.


\begin{thebibliography}{57}
\expandafter\ifx\csname natexlab\endcsname\relax\def\natexlab#1{#1}\fi

\bibitem[{Bahcall} et~al.(1997){Bahcall}, {Kirhakos}, {Saxe}, and
  {Schneider}]{bahcall1997}
{Bahcall}, J.~N., S.~{Kirhakos}, D.~H. {Saxe}, and D.~P. {Schneider}, 1997:
  {Hubble Space Telescope Images of a Sample of 20 Nearby Luminous Quasars}.
  {\em \apj\/}, {\bf 479}, 642.

\bibitem[{Bennert} et~al.(2008){Bennert}, {Canalizo}, {Jungwiert}, {Stockton},
  {Schweizer}, {Peng}, and {Lacy}]{bennert2008}
{Bennert}, N., G.~{Canalizo}, B.~{Jungwiert}, A.~{Stockton}, F.~{Schweizer},
  C.~Y. {Peng}, and M.~{Lacy}, 2008: {Evidence for Merger Remnants in
  Early-Type Host Galaxies of Low-Redshift QSOs}. {\em \apj\/}, {\bf 677},
  846--857.

\bibitem[{Brotherton} et~al.(2002){Brotherton}, {Grabelsky}, {Canalizo}, {van
  Breugel}, {Filippenko}, {Croom}, {Boyle}, and {Shanks}]{brotherton2002}
{Brotherton}, M.~S., M.~{Grabelsky}, G.~{Canalizo}, W.~{van Breugel}, A.~V.
  {Filippenko}, S.~{Croom}, B.~{Boyle}, and T.~{Shanks}, 2002: {Hubble Space
  Telescope Imaging of the Poststarburst Quasar UN J1025-0040: Evidence for
  Recent Star Formation}. {\em \pasp\/}, {\bf 114}, 593--601.

\bibitem[{Bruzual} and {Charlot}(2003)]{bc2003}
{Bruzual}, G. and S.~{Charlot}, 2003: {Stellar population synthesis at the
  resolution of 2003}. {\em \mnras\/}, {\bf 344}, 1000--1028.

\bibitem[{Cales} et~al.(2011){Cales}, {Brotherton}, {Shang}, {Bennert},
  {Canalizo}, {Stoll}, {Ganguly}, {Vanden Berk}, {Paul}, and
  {Diamond-Stanic}]{cales2011}
{Cales}, S.~L., M.~S. {Brotherton}, Z.~{Shang}, V.~N. {Bennert}, G.~{Canalizo},
  R.~{Stoll}, R.~{Ganguly}, D.~{Vanden Berk}, C.~{Paul}, and
  A.~{Diamond-Stanic}, 2011: {Hubble Space Telescope Imaging of Post-starburst
  Quasars}. {\em \apj\/}, {\bf 741}, 106.

\bibitem[{Cales} et~al.(2013){Cales}, {Brotherton}, {Shang}, {Runnoe},
  {DiPompeo}, {Bennert}, {Canalizo}, {Hiner}, {Stoll}, {Ganguly}, and
  {Diamond-Stanic}]{cales2013}
{Cales}, S.~L., M.~S. {Brotherton}, Z.~{Shang}, J.~C. {Runnoe}, M.~A.
  {DiPompeo}, V.~N. {Bennert}, G.~{Canalizo}, K.~D. {Hiner}, R.~{Stoll},
  R.~{Ganguly}, and A.~{Diamond-Stanic}, 2013: {The Properties of
  Post-starburst Quasars Based on Optical Spectroscopy}. {\em \apj\/}, {\bf
  762}, 90.

\bibitem[{Canalizo} et~al.(2007){Canalizo}, {Bennert}, {Jungwiert}, {Stockton},
  {Schweizer}, {Lacy}, and {Peng}]{canalizo2007}
{Canalizo}, G., N.~{Bennert}, B.~{Jungwiert}, A.~{Stockton}, F.~{Schweizer},
  M.~{Lacy}, and C.~{Peng}, 2007: {Spectacular Shells in the Host Galaxy of the
  QSO MC2 1635+119}. {\em \apj\/}, {\bf 669}, 801--809.

\bibitem[{Canalizo} and {Stockton}(1997)]{canalizo1997}
{Canalizo}, G. and A.~{Stockton}, 1997: {Spectroscopy of Close Companions to
  Quasi-Stellar Objects and the Ages of Interaction-Induced Starbursts}. {\em
  \apjl\/}, {\bf 480}, L5.

\bibitem[{Canalizo} and {Stockton}(2000{\natexlab{a}})]{canalizo2000a}
---, 2000{\natexlab{a}}: {3C 48: Stellar Populations and the Kinematics of
  Stars and Gas in the Host Galaxy}. {\em \apj\/}, {\bf 528}, 201--218.

\bibitem[{Canalizo} and {Stockton}(2000{\natexlab{b}})]{canalizo2000b}
---, 2000{\natexlab{b}}: {Stellar Populations in the Host Galaxies of Markarian
  1014, IRAS 07598+6508, and Markarian 231}. {\em \aj\/}, {\bf 120},
  1750--1763.

\bibitem[{Canalizo} and {Stockton}(2001)]{canalizo2001}
---, 2001: {Quasi-Stellar Objects, Ultraluminous Infrared Galaxies, and
  Mergers}. {\em \apj\/}, {\bf 555}, 719--743.

\bibitem[{Cappellari} and {Emsellem}(2004)]{cappellari2004}
{Cappellari}, M. and E.~{Emsellem}, 2004: {Parametric Recovery of Line-of-Sight
  Velocity Distributions from Absorption-Line Spectra of Galaxies via Penalized
  Likelihood}. {\em \pasp\/}, {\bf 116}, 138--147.

\bibitem[{Cid Fernandes} et~al.(2005){Cid Fernandes}, {Mateus}, {Sodr{\'e}},
  {Stasi{\'n}ska}, and {Gomes}]{cidfernandes2005}
{Cid Fernandes}, R., A.~{Mateus}, L.~{Sodr{\'e}}, G.~{Stasi{\'n}ska}, and J.~M.
  {Gomes}, 2005: {Semi-empirical analysis of Sloan Digital Sky Survey galaxies
  - I. Spectral synthesis method}. {\em \mnras\/}, {\bf 358}, 363--378.

\bibitem[{Cisternas} et~al.(2011){Cisternas}, {Jahnke}, {Inskip}, {Kartaltepe},
  {Koekemoer}, {Lisker}, {Robaina}, {Scodeggio}, {Sheth}, {Trump}, {Andrae},
  {Miyaji}, {Lusso}, {Brusa}, {Capak}, {Cappelluti}, {Civano}, {Ilbert},
  {Impey}, {Leauthaud}, {Lilly}, {Salvato}, {Scoville}, and
  {Taniguchi}]{cisternas2011}
{Cisternas}, M., K.~{Jahnke}, K.~J. {Inskip}, J.~{Kartaltepe}, A.~M.
  {Koekemoer}, T.~{Lisker}, A.~R. {Robaina}, M.~{Scodeggio}, K.~{Sheth}, J.~R.
  {Trump}, R.~{Andrae}, T.~{Miyaji}, E.~{Lusso}, M.~{Brusa}, P.~{Capak},
  N.~{Cappelluti}, F.~{Civano}, O.~{Ilbert}, C.~D. {Impey}, A.~{Leauthaud},
  S.~J. {Lilly}, M.~{Salvato}, N.~Z. {Scoville}, and Y.~{Taniguchi}, 2011: {The
  Bulk of the Black Hole Growth Since z \~{} 1 Occurs in a Secular Universe: No
  Major Merger-AGN Connection}. {\em \apj\/}, {\bf 726}, 57.

\bibitem[{Connolly} et~al.(1995){Connolly}, {Szalay}, {Bershady}, {Kinney}, and
  {Calzetti}]{connolly1995}
{Connolly}, A.~J., A.~S. {Szalay}, M.~A. {Bershady}, A.~L. {Kinney}, and
  D.~{Calzetti}, 1995: {Spectral Classification of Galaxies: an Orthogonal
  Approach}. {\em \aj\/}, {\bf 110}, 1071.

\bibitem[{Cox} et~al.(2008){Cox}, {Jonsson}, {Somerville}, {Primack}, and
  {Dekel}]{cox2008}
{Cox}, T.~J., P.~{Jonsson}, R.~S. {Somerville}, J.~R. {Primack}, and
  A.~{Dekel}, 2008: {The effect of galaxy mass ratio on merger-driven
  starbursts}. {\em \mnras\/}, {\bf 384}, 386--409.

\bibitem[{Decarli} et~al.(2010){Decarli}, {Falomo}, {Treves}, {Labita},
  {Kotilainen}, and {Scarpa}]{decarli2010}
{Decarli}, R., R.~{Falomo}, A.~{Treves}, M.~{Labita}, J.~K. {Kotilainen}, and
  R.~{Scarpa}, 2010: {The quasar MBH-Mhost relation through cosmic time - II.
  Evidence for evolution from z = 3 to the present age}. {\em \mnras\/}, {\bf
  402}, 2453--2461.

\bibitem[{Di Matteo} et~al.(2007){Di Matteo}, {Combes}, {Melchior}, and
  {Semelin}]{dimatteo2007}
{Di Matteo}, P., F.~{Combes}, A.-L. {Melchior}, and B.~{Semelin}, 2007: {Star
  formation efficiency in galaxy interactions and mergers: a statistical
  study}. {\em \aap\/}, {\bf 468}, 61--81.

\bibitem[{Di Matteo} et~al.(2005){Di Matteo}, {Springel}, and
  {Hernquist}]{dimatteo2005}
{Di Matteo}, T., V.~{Springel}, and L.~{Hernquist}, 2005: {Energy input from
  quasars regulates the growth and activity of black holes and their host
  galaxies}. {\em \nat\/}, {\bf 433}, 604--607.

\bibitem[{Dunlop} et~al.(2003){Dunlop}, {McLure}, {Kukula}, {Baum}, {O'Dea},
  and {Hughes}]{dunlop2003}
{Dunlop}, J.~S., R.~J. {McLure}, M.~J. {Kukula}, S.~A. {Baum}, C.~P. {O'Dea},
  and D.~H. {Hughes}, 2003: {Quasars, their host galaxies and their central
  black holes}. {\em \mnras\/}, {\bf 340}, 1095--1135.

\bibitem[{Dunlop} et~al.(1993){Dunlop}, {Taylor}, {Hughes}, and
  {Robson}]{dunlop1993}
{Dunlop}, J.~S., G.~L. {Taylor}, D.~H. {Hughes}, and E.~I. {Robson}, 1993:
  {Infrared Imaging of the Host Galaxies of Radio-Loud and Radio-Quiet
  Quasars}. {\em \mnras\/}, {\bf 264}, 455.

\bibitem[{Ferrarese} and {Merritt}(2000)]{fm2000}
{Ferrarese}, L. and D.~{Merritt}, 2000: {A Fundamental Relation between
  Supermassive Black Holes and Their Host Galaxies}. {\em \apjl\/}, {\bf 539},
  L9--L12.

\bibitem[{Fu} and {Stockton}(2009)]{fu2009}
{Fu}, H. and A.~{Stockton}, 2009: {FR II Quasars: Infrared Properties, Star
  Formation Rates, and Extended Ionized Gas}. {\em \apj\/}, {\bf 696},
  1693--1699.

\bibitem[{Gebhardt} et~al.(2000){Gebhardt}, {Bender}, {Bower}, {Dressler},
  {Faber}, {Filippenko}, {Green}, {Grillmair}, {Ho}, {Kormendy}, {Lauer},
  {Magorrian}, {Pinkney}, {Richstone}, and {Tremaine}]{geb00a}
{Gebhardt}, K., R.~{Bender}, G.~{Bower}, A.~{Dressler}, S.~M. {Faber}, A.~V.
  {Filippenko}, R.~{Green}, C.~{Grillmair}, L.~C. {Ho}, J.~{Kormendy}, T.~R.
  {Lauer}, J.~{Magorrian}, J.~{Pinkney}, D.~{Richstone}, and S.~{Tremaine},
  2000: {A Relationship between Nuclear Black Hole Mass and Galaxy Velocity
  Dispersion}. {\em \apjl\/}, {\bf 539}, L13--L16.

\bibitem[{Hopkins}(2012)]{hopkins2012}
{Hopkins}, P.~F., 2012: {Dynamical delays between starburst and AGN activity in
  galaxy nuclei}. {\em \mnras\/}, {\bf 420}, L8--L12.

\bibitem[{Hopkins} et~al.(2005){Hopkins}, {Hernquist}, {Cox}, {Di Matteo},
  {Robertson}, and {Springel}]{hopkins2005a}
{Hopkins}, P.~F., L.~{Hernquist}, T.~J. {Cox}, T.~{Di Matteo}, B.~{Robertson},
  and V.~{Springel}, 2005: {Luminosity-dependent Quasar Lifetimes: A New
  Interpretation of the Quasar Luminosity Function}. {\em \apj\/}, {\bf 630},
  716--720.

\bibitem[{Hopkins} et~al.(2006{\natexlab{a}}){Hopkins}, {Hernquist}, {Cox}, {Di
  Matteo}, {Robertson}, and {Springel}]{hopkins2006a}
---, 2006{\natexlab{a}}: {A Unified, Merger-driven Model of the Origin of
  Starbursts, Quasars, the Cosmic X-Ray Background, Supermassive Black Holes,
  and Galaxy Spheroids}. {\em \apjs\/}, {\bf 163}, 1--49.

\bibitem[{Hopkins} et~al.(2008){Hopkins}, {Hernquist}, {Cox}, and {Kere{\v
  s}}]{hopkins2008}
{Hopkins}, P.~F., L.~{Hernquist}, T.~J. {Cox}, and D.~{Kere{\v s}}, 2008: {A
  Cosmological Framework for the Co-Evolution of Quasars, Supermassive Black
  Holes, and Elliptical Galaxies. I. Galaxy Mergers and Quasar Activity}. {\em
  \apjs\/}, {\bf 175}, 356--389.

\bibitem[{Hopkins} et~al.(2006{\natexlab{b}}){Hopkins}, {Somerville},
  {Hernquist}, {Cox}, {Robertson}, and {Li}]{hopkins2006b}
{Hopkins}, P.~F., R.~S. {Somerville}, L.~{Hernquist}, T.~J. {Cox},
  B.~{Robertson}, and Y.~{Li}, 2006{\natexlab{b}}: {The Relation between Quasar
  and Merging Galaxy Luminosity Functions and the Merger-driven Star Formation
  History of the Universe}. {\em \apj\/}, {\bf 652}, 864--888.

\bibitem[{Hughes} et~al.(2000){Hughes}, {Kukula}, {Dunlop}, and
  {Boroson}]{hughes2000}
{Hughes}, D.~H., M.~J. {Kukula}, J.~S. {Dunlop}, and T.~{Boroson}, 2000:
  {Optical off-nuclear spectra of quasar hosts and radio galaxies}. {\em
  \mnras\/}, {\bf 316}, 204--224.

\bibitem[{Jahnke} et~al.(2004){Jahnke}, {S{\'a}nchez}, {Wisotzki}, {Barden},
  {Beckwith}, {Bell}, {Borch}, {Caldwell}, {H{\"a}ussler}, {Heymans}, {Jogee},
  {McIntosh}, {Meisenheimer}, {Peng}, {Rix}, {Somerville}, and
  {Wolf}]{jahnke2004}
{Jahnke}, K., S.~F. {S{\'a}nchez}, L.~{Wisotzki}, M.~{Barden}, S.~V.~W.
  {Beckwith}, E.~F. {Bell}, A.~{Borch}, J.~A.~R. {Caldwell}, B.~{H{\"a}ussler},
  C.~{Heymans}, S.~{Jogee}, D.~H. {McIntosh}, K.~{Meisenheimer}, C.~Y. {Peng},
  H.~{Rix}, R.~S. {Somerville}, and C.~{Wolf}, 2004: {Ultraviolet Light from
  Young Stars in GEMS Quasar Host Galaxies at $1.8<z<2.75$}. {\em \apj\/}, {\bf
  614}, 568--585.

\bibitem[{Jahnke} et~al.(2007){Jahnke}, {Wisotzki}, {Courbin}, and
  {Letawe}]{jahnke2007}
{Jahnke}, K., L.~{Wisotzki}, F.~{Courbin}, and G.~{Letawe}, 2007: {Spatial
  decomposition of on-nucleus spectra of quasar host galaxies}. {\em \mnras\/},
  {\bf 378}, 23--40.

\bibitem[{Jimenez} et~al.(2004){Jimenez}, {MacDonald}, {Dunlop}, {Padoan}, and
  {Peacock}]{jimenez2004}
{Jimenez}, R., J.~{MacDonald}, J.~S. {Dunlop}, P.~{Padoan}, and J.~A.
  {Peacock}, 2004: {Synthetic stellar populations: single stellar populations,
  stellar interior models and primordial protogalaxies}. {\em \mnras\/}, {\bf
  349}, 240--254.

\bibitem[{Kauffmann} et~al.(2003){Kauffmann}, {Heckman}, {Tremonti},
  {Brinchmann}, {Charlot}, {White}, {Ridgway}, {Brinkmann}, {Fukugita}, {Hall},
  {Ivezi{\'c}}, {Richards}, and {Schneider}]{kauffmann2003}
{Kauffmann}, G., T.~M. {Heckman}, C.~{Tremonti}, J.~{Brinchmann}, S.~{Charlot},
  S.~D.~M. {White}, S.~E. {Ridgway}, J.~{Brinkmann}, M.~{Fukugita}, P.~B.
  {Hall}, {\v Z}.~{Ivezi{\'c}}, G.~T. {Richards}, and D.~P. {Schneider}, 2003:
  {The host galaxies of active galactic nuclei}. {\em \mnras\/}, {\bf 346},
  1055--1077.

\bibitem[{Kocevski} et~al.(2012){Kocevski}, {Faber}, {Mozena}, {Koekemoer},
  {Nandra}, {Rangel}, {Laird}, {Brusa}, {Wuyts}, {Trump}, {Koo}, {Somerville},
  {Bell}, {Lotz}, {Alexander}, {Bournaud}, {Conselice}, {Dahlen}, {Dekel},
  {Donley}, {Dunlop}, {Finoguenov}, {Georgakakis}, {Giavalisco}, {Guo},
  {Grogin}, {Hathi}, {Juneau}, {Kartaltepe}, {Lucas}, {McGrath}, {McIntosh},
  {Mobasher}, {Robaina}, {Rosario}, {Straughn}, {van der Wel}, and
  {Villforth}]{kocevski2012}
{Kocevski}, D.~D., S.~M. {Faber}, M.~{Mozena}, A.~M. {Koekemoer}, K.~{Nandra},
  C.~{Rangel}, E.~S. {Laird}, M.~{Brusa}, S.~{Wuyts}, J.~R. {Trump}, D.~C.
  {Koo}, R.~S. {Somerville}, E.~F. {Bell}, J.~M. {Lotz}, D.~M. {Alexander},
  F.~{Bournaud}, C.~J. {Conselice}, T.~{Dahlen}, A.~{Dekel}, J.~L. {Donley},
  J.~S. {Dunlop}, A.~{Finoguenov}, A.~{Georgakakis}, M.~{Giavalisco}, Y.~{Guo},
  N.~A. {Grogin}, N.~P. {Hathi}, S.~{Juneau}, J.~S. {Kartaltepe}, R.~A.
  {Lucas}, E.~J. {McGrath}, D.~H. {McIntosh}, B.~{Mobasher}, A.~R. {Robaina},
  D.~{Rosario}, A.~N. {Straughn}, A.~{van der Wel}, and C.~{Villforth}, 2012:
  {CANDELS: Constraining the AGN-Merger Connection with Host Morphologies at z
  \~{} 2}. {\em \apj\/}, {\bf 744}, 148.

\bibitem[{Letawe} et~al.(2007){Letawe}, {Magain}, {Courbin}, {Jablonka},
  {Jahnke}, {Meylan}, and {Wisotzki}]{letawe2007}
{Letawe}, G., P.~{Magain}, F.~{Courbin}, P.~{Jablonka}, K.~{Jahnke},
  G.~{Meylan}, and L.~{Wisotzki}, 2007: {On-axis spectroscopy of the host
  galaxies of 20 optically luminous quasars at z \~{} 0.3}. {\em \mnras\/},
  {\bf 378}, 83--108.

\bibitem[{Maraston}(2005)]{maraston2005}
{Maraston}, C., 2005: {Evolutionary population synthesis: models, analysis of
  the ingredients and application to high-z galaxies}. {\em \mnras\/}, {\bf
  362}, 799--825.

\bibitem[{Massey} et~al.(1988){Massey}, {Strobel}, {Barnes}, and
  {Anderson}]{massey1988}
{Massey}, P., K.~{Strobel}, J.~V. {Barnes}, and E.~{Anderson}, 1988:
  {Spectrophotometric standards}. {\em \apj\/}, {\bf 328}, 315--333.

\bibitem[{McLure} et~al.(1999){McLure}, {Kukula}, {Dunlop}, {Baum}, {O'Dea},
  and {Hughes}]{mclure1999}
{McLure}, R.~J., M.~J. {Kukula}, J.~S. {Dunlop}, S.~A. {Baum}, C.~P. {O'Dea},
  and D.~H. {Hughes}, 1999: {A comparative HST imaging study of the host
  galaxies of radio-quiet quasars, radio-loud quasars and radio galaxies - I}.
  {\em \mnras\/}, {\bf 308}, 377--404.

\bibitem[{Mihos} and {Hernquist}(1996)]{mihos1996}
{Mihos}, J.~C. and L.~{Hernquist}, 1996: {Gasdynamics and Starbursts in Major
  Mergers}. {\em \apj\/}, {\bf 464}, 641.

\bibitem[{Nolan} et~al.(2001){Nolan}, {Dunlop}, {Kukula}, {Hughes}, {Boroson},
  and {Jimenez}]{nolan2001}
{Nolan}, L.~A., J.~S. {Dunlop}, M.~J. {Kukula}, D.~H. {Hughes}, T.~{Boroson},
  and R.~{Jimenez}, 2001: {The ages of quasar host galaxies}. {\em \mnras\/},
  {\bf 323}, 308--330.

\bibitem[{Oke} et~al.(1995){Oke}, {Cohen}, {Carr}, {Cromer}, {Dingizian},
  {Harris}, {Labrecque}, {Lucinio}, {Schaal}, {Epps}, and {Miller}]{oke1995}
{Oke}, J.~B., J.~G. {Cohen}, M.~{Carr}, J.~{Cromer}, A.~{Dingizian}, F.~H.
  {Harris}, S.~{Labrecque}, R.~{Lucinio}, W.~{Schaal}, H.~{Epps}, and
  J.~{Miller}, 1995: {The Keck Low-Resolution Imaging Spectrometer}. {\em
  \pasp\/}, {\bf 107}, 375.

\bibitem[{S{\'a}nchez} et~al.(2004){S{\'a}nchez}, {Jahnke}, {Wisotzki},
  {McIntosh}, {Bell}, {Barden}, {Beckwith}, {Borch}, {Caldwell},
  {H{\"a}ussler}, {Jogee}, {Meisenheimer}, {Peng}, {Rix}, {Somerville}, and
  {Wolf}]{sanchez2004}
{S{\'a}nchez}, S.~F., K.~{Jahnke}, L.~{Wisotzki}, D.~H. {McIntosh}, E.~F.
  {Bell}, M.~{Barden}, S.~V.~W. {Beckwith}, A.~{Borch}, J.~A.~R. {Caldwell},
  B.~{H{\"a}ussler}, S.~{Jogee}, K.~{Meisenheimer}, C.~Y. {Peng}, H.~{Rix},
  R.~S. {Somerville}, and C.~{Wolf}, 2004: {Colors of Active Galactic Nucleus
  Host Galaxies at $0.5<z<1.1$ from the GEMS Survey}. {\em \apj\/}, {\bf 614},
  586--606.

\bibitem[{Schlegel} et~al.(1998){Schlegel}, {Finkbeiner}, and
  {Davis}]{schlegel1998}
{Schlegel}, D.~J., D.~P. {Finkbeiner}, and M.~{Davis}, 1998: {Maps of Dust
  Infrared Emission for Use in Estimation of Reddening and Cosmic Microwave
  Background Radiation Foregrounds}. {\em \apj\/}, {\bf 500}, 525--+.

\bibitem[{Schmidt} and {Green}(1983)]{schmidt1983}
{Schmidt}, M. and R.~F. {Green}, 1983: {Quasar evolution derived from the
  Palomar bright quasar survey and other complete quasar surveys}. {\em
  \apj\/}, {\bf 269}, 352--374.

\bibitem[{Shi} et~al.(2007){Shi}, {Ogle}, {Rieke}, {Antonucci}, {Hines},
  {Smith}, {Low}, {Bouwman}, and {Willmer}]{shi2007}
{Shi}, Y., P.~{Ogle}, G.~H. {Rieke}, R.~{Antonucci}, D.~C. {Hines}, P.~S.
  {Smith}, F.~J. {Low}, J.~{Bouwman}, and C.~{Willmer}, 2007: {Aromatic
  Features in AGNs: Star-forming Infrared Luminosity Function of AGN Host
  Galaxies}. {\em \apj\/}, {\bf 669}, 841--861.

\bibitem[{Shi} et~al.(2009){Shi}, {Rieke}, {Ogle}, {Jiang}, and
  {Diamond-Stanic}]{shi2009}
{Shi}, Y., G.~H. {Rieke}, P.~{Ogle}, L.~{Jiang}, and A.~M. {Diamond-Stanic},
  2009: {Cosmic Evolution of Star Formation in Type-1 Quasar Hosts Since z =
  1}. {\em \apj\/}, {\bf 703}, 1107--1122.

\bibitem[{Spergel} et~al.(2003){Spergel}, {Verde}, {Peiris}, {Komatsu},
  {Nolta}, {Bennett}, {Halpern}, {Hinshaw}, {Jarosik}, {Kogut}, {Limon},
  {Meyer}, {Page}, {Tucker}, {Weiland}, {Wollack}, and {Wright}]{spergel2003}
{Spergel}, D.~N., L.~{Verde}, H.~V. {Peiris}, E.~{Komatsu}, M.~R. {Nolta},
  C.~L. {Bennett}, M.~{Halpern}, G.~{Hinshaw}, N.~{Jarosik}, A.~{Kogut},
  M.~{Limon}, S.~S. {Meyer}, L.~{Page}, G.~S. {Tucker}, J.~L. {Weiland},
  E.~{Wollack}, and E.~L. {Wright}, 2003: {First-Year Wilkinson Microwave
  Anisotropy Probe (WMAP) Observations: Determination of Cosmological
  Parameters}. {\em \apjs\/}, {\bf 148}, 175--194.

\bibitem[{Stockton} and {MacKenty}(1987)]{stockton1987}
{Stockton}, A. and J.~W. {MacKenty}, 1987: {Extended emission-line regions
  around QSOs}. {\em \apj\/}, {\bf 316}, 584--596.

\bibitem[{Torrey} et~al.(2012){Torrey}, {Cox}, {Kewley}, and
  {Hernquist}]{torrey2012}
{Torrey}, P., T.~J. {Cox}, L.~{Kewley}, and L.~{Hernquist}, 2012: {The
  Metallicity Evolution of Interacting Galaxies}. {\em \apj\/}, {\bf 746}, 108.

\bibitem[{Trager} et~al.(2005){Trager}, {Worthey}, {Faber}, and
  {Dressler}]{trager2005}
{Trager}, S.~C., G.~{Worthey}, S.~M. {Faber}, and A.~{Dressler}, 2005: {Hot
  stars in old stellar populations: a continuing need for intermediate ages}.
  {\em \mnras\/}, {\bf 362}, 2--8.

\bibitem[{Treister} et~al.(2012){Treister}, {Schawinski}, {Urry}, and
  {Simmons}]{treister2012}
{Treister}, E., K.~{Schawinski}, C.~M. {Urry}, and B.~D. {Simmons}, 2012:
  {Major Galaxy Mergers Only Trigger the Most Luminous Active Galactic Nuclei}.
  {\em \apjl\/}, {\bf 758}, L39.

\bibitem[{Van Wassenhove} et~al.(2012){Van Wassenhove}, {Volonteri}, {Mayer},
  {Dotti}, {Bellovary}, and {Callegari}]{vanwassenhove2012}
{Van Wassenhove}, S., M.~{Volonteri}, L.~{Mayer}, M.~{Dotti}, J.~{Bellovary},
  and S.~{Callegari}, 2012: {Observability of Dual Active Galactic Nuclei in
  Merging Galaxies}. {\em \apjl\/}, {\bf 748}, L7.

\bibitem[{Vanden Berk} et~al.(2006){Vanden Berk}, {Shen}, {Yip}, {Schneider},
  {Connolly}, {Burton}, {Jester}, {Hall}, {Szalay}, and
  {Brinkmann}]{vandenberk2006}
{Vanden Berk}, D.~E., J.~{Shen}, C.-W. {Yip}, D.~P. {Schneider}, A.~J.
  {Connolly}, R.~E. {Burton}, S.~{Jester}, P.~B. {Hall}, A.~S. {Szalay}, and
  J.~{Brinkmann}, 2006: {Spectral Decomposition of Broad-Line AGNs and Host
  Galaxies}. {\em \aj\/}, {\bf 131}, 84--99.

\bibitem[{Wold} et~al.(2010){Wold}, {Sheinis}, {Wolf}, and {Hooper}]{wold2010}
{Wold}, I., A.~I. {Sheinis}, M.~J. {Wolf}, and E.~J. {Hooper}, 2010: {Host
  galaxies of luminous quasars: population synthesis of optical off-axis
  spectra}. {\em \mnras\/}, {\bf 408}, 713--730.

\bibitem[{Young} et~al.(2009){Young}, {Axon}, {Robinson}, and
  {Capetti}]{young2009}
{Young}, S., D.~J. {Axon}, A.~{Robinson}, and A.~{Capetti}, 2009: {The
  Contribution from Scattered Light to Quasar Galaxy Hosts}. {\em \apjl\/},
  {\bf 698}, L121--L124.

\bibitem[{Zibetti} et~al.(2013){Zibetti}, {Gallazzi}, {Charlot}, {Pierini}, and
  {Pasquali}]{zibetti2013}
{Zibetti}, S., A.~{Gallazzi}, S.~{Charlot}, D.~{Pierini}, and A.~{Pasquali},
  2013: {Near-infrared spectroscopy of post-starburst galaxies: a limited
  impact of TP-AGB stars on galaxy spectral energy distributions}. {\em
  \mnras\/}, {\bf 428}, 1479--1497.

\end{thebibliography}

\clearpage

\end{document}